\documentclass[pdflatex,sn-mathphys]{sn-jnl}

\PassOptionsToPackage{dvipsnames,svgnames,table}{xcolor}
\def\BibTeX{{\rm B\kern-.05em{\sc i\kern-.025em b}\kern-.08em
    T\kern-.1667em\lower.7ex\hbox{E}\kern-.125emX}}

\usepackage[labeled, resetlabels]{multibib}
\newcites{S}{Primary studies}
\usepackage[english]{babel}

\usepackage{longtable}
\usepackage{lscape}
\usepackage{colortbl}
\usepackage{tabulary}
\usepackage{etoolbox}

\UseRawInputEncoding

\usepackage{pifont}

\usepackage{import}
\usepackage{ifxetex,ifluatex}

\usepackage{enumitem}
\newlist{nobullets}{itemize}{1}
\setlist[nobullets,1]{label={}, left=0pt}

\jyear{2021}%

\theoremstyle{thmstyleone}%
\theoremstyle{thmstyletwo}%

\theoremstyle{thmstylethree}%

\begin{document}

\title[RE4AI: A Systematic Mapping Study]{How Mature is Requirements Engineering for AI-based Systems? A Systematic Mapping Study on Practices, Challenges, and Future Research Directions}


\author*[1,2]{\fnm{Umm-e-}\sur{Habiba}}\email{umm-e-habiba@iste.uni-stuttgart.de}

 \author[1]{\fnm{Markus} \sur{Haug}}\email{markus.haug@iste.uni-stuttgart.de}

\author[3]{\fnm{Justus} \sur{Bogner}}\email{j.bogner@vu.nl}

\author[4]{\fnm{Stefan} \sur{Wagner}}\email{stefan.wagner@tum.de}

\affil[1]{\orgdiv{Institute of Software Engineering}, \orgname{University of Stuttgart}, \orgaddress{\street{} \city{} \postcode{70569}, \state{Stuttgart}, \country{Germany}}}

\affil[2]{\orgdiv{Department of Software Engineering}, \orgname{University of Kotli Azad Jammu and Kashmir}, \city{Kotli}, \state{Azad Kashmir}, \country{Pakistan}}

\affil[3]{\orgdiv{Department of Computer Science}, \orgname{Vrije Universiteit Amsterdam}, \city{Amsterdam}, \country{The Netherlands}}

\affil[4]{\orgname{TUM School of Communication, Information and Technology, Technical University of Munich}, \orgaddress{\city{Heilbronn}, \country{Germany}}}


\abstract{
Artificial intelligence (AI) permeates all fields of life, which resulted in new challenges in requirements engineering for artificial intelligence (RE4AI), e.g., the difficulty in specifying and validating requirements for AI or considering new quality requirements due to emerging ethical implications. It is currently unclear if existing RE methods are sufficient or if new ones are needed to address these challenges.
Therefore, our goal is to provide a comprehensive overview of RE4AI to researchers and practitioners. What has been achieved so far, i.e., what practices are available, and what research gaps and challenges still need to be addressed? 
To achieve this, we conducted a systematic mapping study combining query string search and extensive snowballing. The extracted data was aggregated, and results were synthesized using thematic analysis.
Our selection process led to the inclusion of 126 primary studies.
Existing RE4AI research focuses mainly on requirements analysis and elicitation, with most practices applied in these areas.
Furthermore, we identified requirements specification, explainability, and the gap between machine learning engineers and end-users as the most prevalent challenges, along with a few others. Additionally, we proposed seven potential research directions to address these challenges.
Practitioners can use our results to identify and select suitable RE methods for working on their AI-based systems, while researchers can build on the identified gaps and research directions to push the field forward. \footnote{This paper has been accepted in Requirements Engineering Journal}

}





\maketitle

\section{Introduction}
\label{Sec:Intro}
In the past decade, the need for automation and intelligence has led to huge advancements in machine learning (ML) and artificial intelligence (AI)~\cite{8474484}.
Despite substantial growth in AI-based systems, we continue to see significant challenges and project failures~\cite{khomh2018software}. Weiner~\cite{weiner2020ai} stated that according to recent data, 87\% of AI projects do not make it into production, meaning that most projects are never deployed.
One primary reason is that AI has disrupted traditional software development practices, which are typically deductive, where requirements are explicitly defined and translated into code. In contrast, AI-based systems, i.e., systems that incorporate AI components~\cite{martinez2022software}, are developed inductively, as they learn and adapt from training data. This shift in approach makes it challenging to anticipate and understand the behavior of AI-based systems. Due to the critical role of AI-based systems, software engineering (SE) and the AI community must collaborate to develop new strategies to address these issues.

Requirements engineering (RE), the systematic handling of software requirements~\cite{weisz2021perfection}, has been impacted by the complexities of AI-based systems~\cite{8933653}. Existing RE approaches are challenging to apply to AI-based systems because of their probabilistic nature and the need for constant adaptation. To address these challenges, RE needs to evolve to be compatible with AI-based systems~\cite{lwakatare2019taxonomy}. 
The roles and responsibilities related to RE have changed, with data scientists now being responsible for specifying high-level requirements in ML systems, which can lead to systems that prioritize data quality over stakeholder requirements~\cite{vogelsang2019requirements}. Hence, software engineers and data scientists must work together to address issues arising from the combination of AI and SE~\cite{amershi2019software}. 
Kondermann~\cite{kondermann2013ground} claimed that despite its long history, RE has yet to be used extensively for AI, especially computer vision, and called for more research to integrate data selection approaches. Further, he emphasized that AI system requirements are complex to elicit, specify, and manage.  

Recently, substantial growth in studies addressing Requirements Engineering for Artificial Intelligence (RE4AI) can be seen. With this proliferation, there is a need to understand what has been achieved so far. This is important for practitioners to identify suitable methods for their day-to-day work as well as for researchers to tackle important challenges.
Therefore, we conducted a systematic mapping study to explore the potential of RE for contributing to AI-based systems.
We chose a systematic mapping study (SMS) over a systematic literature review because our objective was to capture the landscape of research in our area of study, not just to answer specific questions but to understand the diversity and scope of research that has been conducted. An SMS allows us to achieve this by categorizing research works based on various dimensions such as methodology, themes, outcomes, and geographical focus.
This study aims to explore the current RE4AI research landscape regarding RE practices for AI-based systems, topics covered regarding SWEBOK~\cite{bourque2014swebok} areas, maturity of the research area, challenges, and future directions.

The main contributions of this SMS are summarised as follows:
\begin{itemize}

\item We provide a research overview regarding RE4AI, i.e., which topics have been explored according to SWEBOK \cite{bourque2014swebok} and the type of research conducted according to the classification of Wieringa et al.~\cite{wieringa2006requirements}. 

\item We identify which existing RE practices have been applied to AI-based systems and also present new RE practices proposed specifically for AI-based systems.

\item We identify the current trends and challenges in applying the existing RE approaches for AI-based systems.

\item Finally, we extracted potential future research directions from selected primary studies.

\end{itemize}

\textbf{Article Organization:}
The remainder of the article is organized as follows: Section~\ref{sec:RelatedWork} compares existing studies similar to ours. Section~\ref{sec:Preliminaries} describes the research design we followed in this study. Then, Section~\ref{sec:Results} presents the extractions, synthesis, and results from the selected primary studies. Section~\ref{sec:challenges} identifies challenges and future directions whereas, Section~\ref{sec:discussion} discusses our results by providing combined analysis of RQ4 and RQ5, and Section~\ref{sec: Threats to validity} presents threats to validity and how we mitigated them. Finally, Section~\ref{sec:conclusion} concludes this paper.

\section{Related Work}
\label{sec:RelatedWork}

With the recent growth of AI-based systems, the number of empirical studies on RE for AI is increasing, with more and more different aspects being investigated. There exist secondary studies that highlighted research in the field of SE for AI-based systems~\cite{martinez2022software, 8933800, nascimento2020software, lorenzoni2021machine, kumeno2019sofware}. However, these studies are not exclusively focused on the RE process but consider it only as a part of SE.
In relation to RE, we identified three secondary studies on RE for AI/ML~\cite{ahmad2021s, 9582561, 9719885} that are closely related to our work. Ahmad et al.~\cite{ahmad2021s} conducted a systematic literature review and identified $27$ primary studies. They identified the notations and modeling languages utilized in the development of AI systems, focusing specifically on the application domains of these AI systems. Villamizar et al.~\cite{9582561} conducted a mapping study incorporating findings from $35$ distinct studies. They identified the contribution of RE aspects to the development of ML-based systems, focusing on RE topics. Further, they emphasized the quality characteristics that are considered during RE for ML-based systems. Yoshioka et al.~\cite{9719885} also conducted a systematic literature review. Their focus was mainly on current techniques and practices of RE for machine learning systems (MLS). They analyzed $32$ studies and mapped a research landscape of RE techniques and practices while identifying research gaps. 

Our research significantly updates and expands upon previous reviews by analyzing 126 primary studies, broadening the scope to capture developments up to and including July 2023. We contribute by organizing current RE practices within SWEBOK Knowledge Areas (KAs) and spotlighting newly introduced practices.

Our work synthesizes challenges and future research directions from these studies, providing an in-depth overview of the evolving trends in the field. Distinguishing our systematic mapping study, we align our categorization with SWEBOK standards and employ Wieringa's~\cite{wieringa2006requirements} classification and Wohlin's~\cite{wohlin2012empirical} evaluation strategy to assess research maturity. This methodology sets our study apart from others, such as the systematic literature reviews by Ahmad et al.~\cite{ahmad2021s} and Yoshioka et al..~\cite{9719885}, which cover studies up to August 2021, and Villamizar et al.~\cite{9582561}’s study, which concludes in December 2020.

Our inclusion of literature up to and including July 2023 and our broader examination of AI-based systems, encompassing the entirety of AI, allows for a comprehensive understanding of RE challenges and practices relevant to AI. By including an additional two years and 100 more papers, our study reflects the significant shift in focus towards safe and responsible AI, as well as the evolving legal requirements of AI, particularly as AI impacts many more aspects of life following the rise of large language models (LLMs).

We critically assess and identify emerging RE practices tailored for AI, addressing the distinctive needs of AI systems compared to traditional approaches. This thorough analysis leads to a detailed compilation of challenges and directions for future research, significantly advancing the discussion on RE for AI.

We compare the related work to our study in Table~\ref{tab:comparision}. 

\begin{table*}[hbt!]
\footnotesize
\caption{Comparison of related studies with our Systematic Mapping Study}
\label{tab:comparision}
\begin{tabular}{llll}
\hline
\rowcolor[HTML]{FFFFFF} 
{\color[HTML]{0D0D0D} \textbf{Study}} &
  {\color[HTML]{0D0D0D} \textbf{\begin{tabular}[c]{@{}l@{}}Type of Study \&\\ Study Period\end{tabular}}} &
  {\color[HTML]{0D0D0D} \textbf{Focus of the Study}} &
  {\color[HTML]{0D0D0D} \textbf{\begin{tabular}[c]{@{}l@{}}Contribution to RE \\ for AI/ML\end{tabular}}} \\ \hline
\rowcolor[HTML]{FFFFFF} 
{\color[HTML]{0D0D0D} \begin{tabular}[c]{@{}l@{}} Ahmad \\ et al. \end{tabular}} &
  {\color[HTML]{0D0D0D} \begin{tabular}[c]{@{}l@{}}Systematic Literature \\ Review August 2020\\ Identified 27 studies\end{tabular}} &
  {\color[HTML]{0D0D0D} \begin{tabular}[c]{@{}l@{}}Explores approaches \\ for documenting requirements \\ in AI systems, identifying \\ tools, techniques, challenges, \\ and limitations.\end{tabular}} &
  {\color[HTML]{0D0D0D} \begin{tabular}[c]{@{}l@{}}Extracted RE notations\\ and modeling languages\end{tabular}} \\ \hline
\rowcolor[HTML]{FFFFFF} 
{\color[HTML]{0D0D0D} \begin{tabular}[c]{@{}l@{}} Villamizar \\ et al. \end{tabular}} &
  {\color[HTML]{0D0D0D} \begin{tabular}[c]{@{}l@{}}Mapping Study\\ December 2020\\ Identified 35 studies\end{tabular}} &
  {\color[HTML]{0D0D0D} \begin{tabular}[c]{@{}l@{}}Characterizes the RE publication \\ landscape for ML systems, \\ detailing contributions like \\ analyses, approaches, and \\ quality models. Categorizes \\ based on RE activities and \\ empirical evaluation strategies.\end{tabular}} &
  {\color[HTML]{0D0D0D} \begin{tabular}[c]{@{}l@{}}Identified RE aspects \\ contributing to \\ ML-based systems\end{tabular}} \\ \hline
\rowcolor[HTML]{FFFFFF} 
{\color[HTML]{0D0D0D} \begin{tabular}[c]{@{}l@{}} Yoshioka \\ et al. \end{tabular}} &
  {\color[HTML]{0D0D0D} \begin{tabular}[c]{@{}l@{}}Systematic Literature\\ Review August 2021\\ Identified 32 studies\end{tabular}} &
  {\color[HTML]{0D0D0D} \begin{tabular}[c]{@{}l@{}}Assesses techniques and practices \\ of RE for ML, classifying \\ literature by year, venue, ML \\ algorithms, and stakeholders. \\ Highlights research gaps \\ and directions.\end{tabular}} &
  {\color[HTML]{0D0D0D} \begin{tabular}[c]{@{}l@{}}Analyzed current RE\\ techniques and practices \\ for ML systems\end{tabular}} \\ \hline
\rowcolor[HTML]{FFFFFF} 
{\color[HTML]{0D0D0D} Our Study} &
  {\color[HTML]{0D0D0D} \begin{tabular}[c]{@{}l@{}}Systematic Mapping \\ Study up to and\\ including July 2023 \\ Identified 126 studies\end{tabular}} &
  {\color[HTML]{0D0D0D} \begin{tabular}[c]{@{}l@{}}Presents literature demographics, \\ classifies RE contributions by \\ SWEBOK KAs, assesses research \\ maturity. Also, Identifies and \\ analyzes both existing and new RE \\ practices for AI systems, distinctly. \\ Lastly, also proposes a taxonomy \\ for challenges and future research \\ derived from primary studies.\end{tabular}} &
  {\color[HTML]{0D0D0D} \begin{tabular}[c]{@{}l@{}}Expanded the scope \\of reviews, identified \\new RE practices, \& \\synthesized challenges \\and directions for AI\end{tabular}} \\ \hline
\end{tabular}
\end{table*}

\section{Research Design}
\label{sec:Preliminaries}
We conducted this systematic mapping study by following the guidelines of Petersen et al.~\cite{petersen2015guidelines}. In the following, we outline our research questions (RQs) and their rationales, how we obtained articles, and how we addressed them systematically. 

\subsection{Research Questions and Rationales}
\label{sec:RQ}
To define the scope of our study, we formulated RQs that highlight the categorization of the literature in a way that can be interesting for scholars and practitioners, while also providing insights into how RE approaches have been used for AI-based system development. 

\textbf{RQ1: Where and when have RE practices for AI-based systems been published?} 

We aim to explore the recent trends in this research area in terms of the publication year of articles, and how active this area has been. For this purpose, we analyze the study distribution by year of publication and the most preferred venues for RE4AI.

\textbf{RQ2: How are the different RE topics distributed within the literature on AI-based systems?}

We intend to use the SWEBOK~\cite{bourque2014swebok} classification scheme to assign RE categories to our primary studies. The most recently published version of SWEBOK V3.0 has $8$ knowledge areas. Using the RE knowledge area and its categories, we analyze frequently studied RE topics, as well as topics that may require further attention.

\textbf{RQ3: What is the maturity of the research in this area?}

To evaluate the maturity of this area, we analyze the primary studies according to the criteria given below:

\begin{itemize}
\item We use Wieringa et al.'s \cite{wieringa2006requirements} classification scheme for RE publications.
\item We scrutinize if RE techniques discussed in the paper have been empirically evaluated \cite{wohlin2012empirical}.
\item If yes, we analyze if the evaluation took place in an industrial environment.
\end{itemize}

\textbf{RQ4: What RE practices have been proposed for or applied to AI-based systems?}

This question aims to identify new RE practices that have been proposed in the context of AI-based systems and how existing RE practices have been used for AI-based systems. We intend to provide a holistic overview of the degree to which RE has contributed to AI. We answer the following sub-questions:
\begin{itemize}
    \item RQ4.1: Which conventional RE practices have been applied to AI-based systems?
    \item RQ4.2: Which new RE practices have been proposed for AI-based systems?

We classify practices into tools, techniques, processes, and models. To make a clear distinction, we define these terms below.

\noindent
 \textit{Technique:}
 A technique is a specific method applied during a part of the procedure, focusing on actions that are observable and measurable by practitioners~\cite{burnham1992approach},~\cite{hofler1983approach}. For example, a common technique in requirements gathering is the use of interviews, where the interviewer follows a structured approach to elicit information from stakeholders. Another example is prototyping, a technique used to quickly create a working model of a system's features to gather feedback. 

\noindent
 \textit{Process:}
 A process is a set of activities to reach a certain goal, which specifies the concrete activity details and the sequence in which they are performed. However, a \textbf{technique} is a specific way to perform one of those steps.

 \noindent
 \textit{Model:}
A model is an abstraction of a system that describes it from a certain perspective. For example, a UML (Unified Modeling Language) Use Case Diagram serves as an example of a model. In the context of RE for AI, models can also include frameworks or theoretical constructs.
 
 \noindent
 \textit{Tool:}
Any software that has been used to support the RE process.
\end{itemize}

\textbf{RQ5: What makes requirement engineering for AI-based systems challenging, and what are future directions to address?}

To answer this question, we analyze AI-based systems and their key characteristics in the targeted studies. This analysis helps us to identify differences between AI-based systems and conventional systems, as well as RE challenges for AI-based systems. Furthermore, we analyze studies that propose future research directions. To address this RQ, we split it into two sub-research questions:
\begin{itemize}
\item RQ5.1: What are the current challenges in the development of AI-based systems?
\item RQ5.2: What are future research directions?
\end{itemize}

  \begin{figure*}[htb]
  	\includegraphics[width=\textwidth]{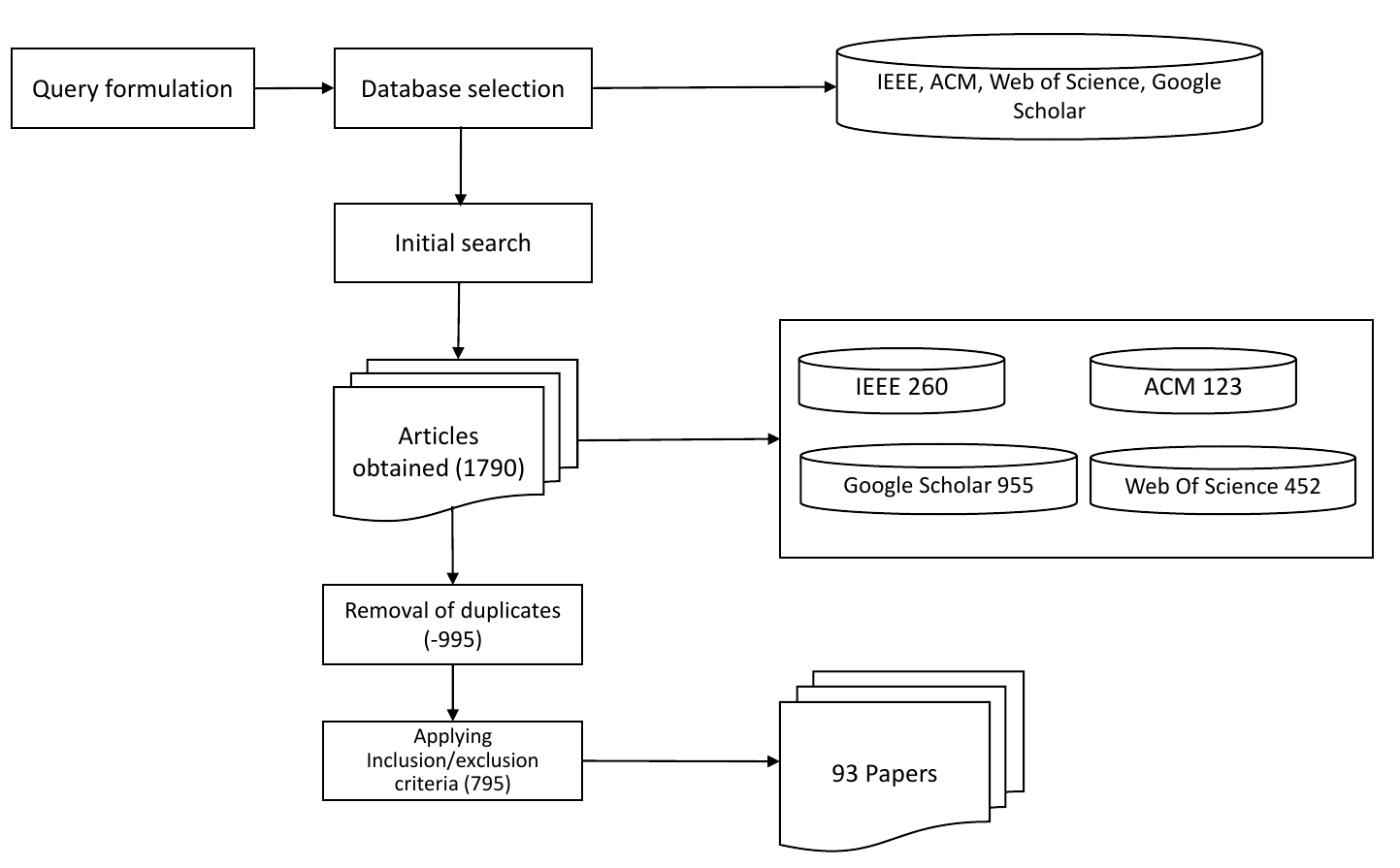}
  	\centering
  	\caption{Search Process}
  	\label{fig:framework}
  	\vspace{-3mm}
  \end{figure*}

\subsection{Research Protocol}

To execute an impartial, objective investigation, a research protocol is required. It manages the flow of research and maximizes the study's valuable findings. We created a research protocol that describes the parts of the study and is depicted in Fig.~\ref{fig:framework}.
The main steps of the research protocol are as follows.

\begin{enumerate}
\item \textbf{Search query formulation:} 

We formulated our search query using the first two elements of the PICO criteria~\cite{petersen2015guidelines}. The first element is the \textit{Population} $(“P”)$, which indicates RE publications.
The second element is \textit{Intervention} $(“I”)$, which specifies AI, where ML and deep learning (DL) are a part of AI. 
We excluded the \textit{Comparison(C)} and \textit{Outcome (O)} criteria from our search to broaden its scope, allowing us to capture a wider range of studies, which is especially useful for exploratory research or obtaining a comprehensive field overview as in an SMS.
We constructed our query iteratively and restricted our search to article titles to achieve the best results. Initially, we evaluated four digital libraries ACM Digital Library, IEEE Xplore, ScienceDirect, and Springer for our search. Our findings indicated that IEEE Xplore and ACM Digital Library were most effective in handling our search query. Concerns may arise regarding the exclusivity of our selection potentially overlooking relevant studies, particularly from Springer. To address this, we incorporated a snowballing technique, systematically reviewing references from our initial findings to ensure no significant work was overlooked. Alongside these selected digital libraries, we expanded our search to include meta-search engines, specifically Google Scholar and Web of Science, focusing on article titles to refine our results. This comprehensive approach, combining direct searches with snowballing, was designed to ensure thorough coverage of the literature.
The finalized search string is given below:

 {\centering \textbf{("requirement" OR "requirements")} \\
\textbf{AND} \\
\textbf{("AI" OR "artificial intelligence" OR "ML" OR "machine learning"
OR "DL" OR "deep learning")}\\}

This query resulted in $123$ articles from ACM Digital Library and $260$ from IEEEXplore, while meta-search engines such as Web of Science returned $452$ and Google Scholar $955$ articles. In total, we obtained $1790$ papers.

\item \textbf{Removal of duplicates:}

In this step, we removed duplicated articles as we ran our query on two digital libraries and two meta-search engines. After the removal of duplicates, we are left with $795$ papers.

\item \textbf{Inclusion/exclusion criteria:}

Our query yielded literature that included all keywords in their title. To rectify the literature according to the scope of our study, we developed inclusion (IC) and exclusion criteria (EC).
\begin{enumerate}
\item \textbf{EC1:} Not relevant to the scope of the study (i.e., studies that do not focus on RE for AI)
\item \textbf{EC2:} Published before $2010$ and after July $2023$
\item \textbf{EC3:} Not written in English
\item \textbf{EC4:} Secondary studies
\item \textbf{EC5:} Not peer-reviewed / not a scientific paper
\item \textbf{EC6:} Not accessible
\item \textbf{IC1:} The primary focus of the paper is requirements engineering
\item \textbf{IC2:} The paper targets AI-based systems
\end{enumerate}

During our study, we used Rayyan~\cite{rayyansystems2022} to remove duplicates and applied inclusion/exclusion of articles. Subsequently, after eliminating duplication, we were left with $795$ unique papers.

The inclusion/exclusion criteria mentioned above were used to refine articles that fit our study scope. EC3 and EC6 were designed to exclude studies that are not written in English and are accessible through any source. At the same time, EC2 is intended to strictly select the publication years considered during the mapping study. To exclude the secondary studies and the grey literature, we apply EC4 and EC5. IC1, IC2, and EC1 required an in-depth study of articles to analyze whether the article fits the scope of the study. 
The first three authors independently assessed $795$ studies using these criteria. Discrepancies in their selections were resolved through discussion and consensus voting. This rigorous process resulted in the selection of $93$ articles.

\item \textbf{Data extraction:}

We extracted data from each primary study to answer our research questions described in section~\ref{sec:RQ} above. We defined extraction sheets to record the necessary information related to each publication. Having a specified extraction sheet will reduce the opportunity to include researcher bias. As a result, during data extraction, the researcher extracted data that should answer the research questions.

Before we started extracting data, a pilot extraction process was conducted to develop a shared understanding and avoid any confusion regarding the extraction process. This pilot ensured that each researcher clearly understood the research questions and respective extraction sheets. For this purpose, we selected three initial studies, and each researcher independently extracted data into their sheet. Afterward, we discussed the extracted data and further improved the extraction sheets. 

We outlined the individual data cells according to each research question. Since each RQ has multiple fields, we maintained a separate spreadsheet for each RQ. RQ1 is primarily focused on studying metadata, including the year of publication, publishing venue, and the involved research community (Note: we classified papers based on author affiliations as industry, academic, or collaborative). To identify which RE topics have been covered frequently within the literature (RQ2), we classified the primary studies using the SWEBOK~\cite{bourque2014swebok} subcategories for RE. Moreover, to judge the maturity of this research area (RQ3), we classified literature according to the RE publication types proposed by Wieringa et al.~\cite{wieringa2006requirements} as well as empirical evaluation method~\cite{wohlin2012empirical}. 

We characterized RE practices in four dimensions: tool, techniques, model, and process. Therefore, RQ4 is designed to capture these details and differentiate between practices that are new or already existing. Finally, we extracted challenges highlighted by different authors and synthesized literature to outline possible research directions (RQ5).

  \begin{figure*}[htb]
  	\includegraphics[width=\textwidth]{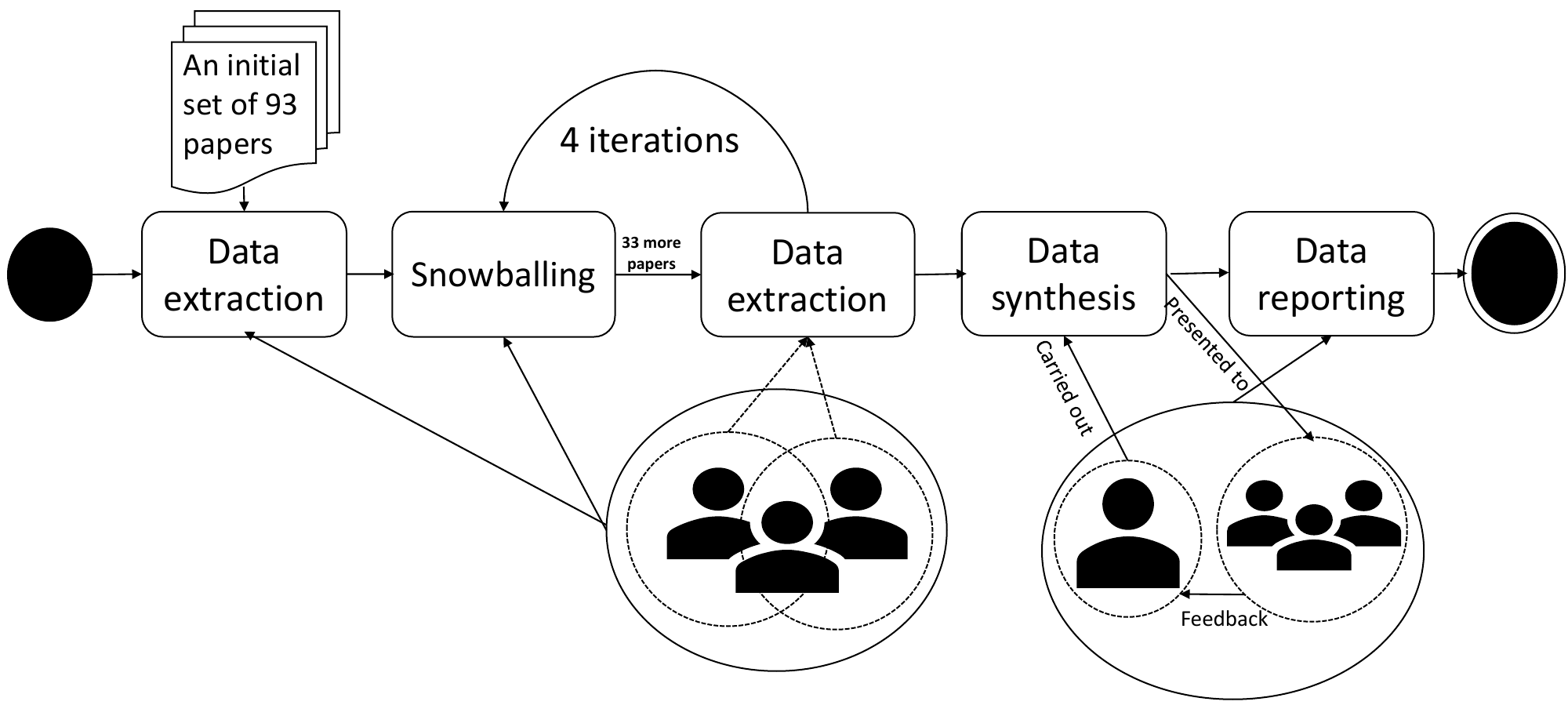}
  	\centering
  	\caption{Data extraction and synthesis process}
  	\label{fig:data-analysis}
  	\vspace{-3mm}
  \end{figure*}

Eventually, we split up the extractions and assigned two researchers to each paper, and after every week, a synchronization meeting was held to discuss extraction as shown in Fig.~\ref{fig:data-analysis}. A separate consensus spreadsheet was maintained where all finalized entries were recorded.
Further, to analyse inter-rater reliability, the agreement level was measured using Cohen's kappa coefficient, which provides a robust statistical measure of inter-rater reliability.

For the inclusion/exclusion criteria, the researchers used the rayyan.ai tool and independently performed the inclusion/exclusion of papers. However, there was no conflict found during this process.

For the thematic analysis, the researchers evaluated 5 papers with 4 questions each, making a total of 20 evaluations.

\begin{itemize}
    \item Total items evaluated (N): 5 papers x 4 questions = 20
    \item Agreement on all 4 questions for 3 papers: 3 x 4 = 12 agreements
    \item Agreement on 3 questions for 1 paper: 3 agreements
    \item Agreement on 2 questions for 1 paper: 2 agreements
    \item Total agreements (A): 12 + 3 + 2 = 17
    \item Total disagreements (D): 20 - 17 = 3
\end{itemize}
\medskip

Cohen's kappa $(\kappa)$ is calculated as follows:

\centerline{$\kappa = \frac{P_o - P_e}{1 - P_e}$}

Where $P_o$ is the observed agreement and $P_e$ is the expected agreement by chance.

\begin{enumerate}
    \item Observed Agreement $P_o$: 
    Number of agreements $/$ Total items evaluated =$17 / 20 = 0.85$
    
    \medskip
    
    \item Expected Agreement $P_e$:
    Assuming equal probability for agreement and disagreement:     
    $P_e$=(Probability of both agreeing)2+(Probability of both disagreeing)
    
    Again, assuming equal distribution (0.5 for agreement and 0.5 for disagreement): 
    
    $P_e$= $(0.5\times 0.5)+(0.5\times0.5)=0.25+0.25=0.50$
    
    \medskip
    
    \item Cohen's Kappa ($\kappa$):
    
    \medskip
    $\kappa = \frac{0.85 - 0.50}{1 - 0.50} = \frac{0.35}{0.50} = 0.70$
\end{enumerate}

\bigskip

These kappa values indicate a substantial level of agreement between the researchers, supporting the reliability of the conclusions derived from their analyses.

\item \textbf{Snowballing:}

Following the first iteration of extractions, we applied forward and backward snowballing according to the guidelines by Wohlin~\cite{wohlin2014guidelines}. Snowballing, which involves using the references of identified papers to find additional relevant studies, can be especially effective in fields where consistent terminology is lacking. This approach helped us identify important studies that might be missed due to inconsistent keyword use in database searches. By using snowballing, we ensured a more comprehensive review by capturing relevant research that might not be easy to find through traditional database searches alone.
To ensure overall coverage, snowballing iterations were performed until no further studies were included.
The first round on the start set of $93$ articles yielded $15$ additional papers. 
After extracting these $15$ papers, the second round of snowballing was carried out, which resulted in $15$ more articles. We then performed a third iteration of snowballing, which yielded $3$ more articles. 
Lastly, snowballing on these $3$ articles did not result in additional papers. 
After this process, we ended with a final set of $126$ primary studies.

\item \textbf{Data synthesis:}

We began data analysis and synthesis once extractions had been completed. To categorize the retrieved data, we used both quantitative and qualitative analysis. Some extraction discrepancies and errors were detected throughout this process and were removed. The first author performed the synthesis and frequently presented the results to the rest, leading to iterative refinements.

To address RQ1, we conducted a frequency analysis to examine bibliographical data. For RQ2 and RQ3, we undertook quantitative analyses. Specifically, the analysis for RQ2 categorized literature based on SWEBOK KAs, whereas RQ3 focused on classifying literature according to Wieringa's~\cite{wieringa2006requirements} framework. Additionally, we identified and analyzed the evaluation methods used for RE practices. To respond to RQ4, we employed a combination of methods, including qualitative analysis through thematic analysis, as recommended by Cruzes et al.~\cite{cruzes2011recommended}.

As RQ5 is divided into two sub-questions, we conducted a qualitative analysis for both questions. Further, we applied the thematic synthesis approach recommended by Cruzes et al.~\cite{cruzes2011recommended}. We then extracted free text from the papers and labeled the free text. Finally, we identified the most recurrent themes in the next step and assigned them to extracted text.
In the following section, we present our data extraction results and their mapping to respective research questions.

\end{enumerate}

\section{Results}
\label{sec:Results}
After data extraction, we move towards the data analysis phase. In this section, we summarise the results of our mapping study. Starting from RQ1, we systematically present results for each RQ, respectively.

\subsection{Bibliometrics (RQ1)}

\begin{figure}
 \centering
 \includegraphics[width=\textwidth]{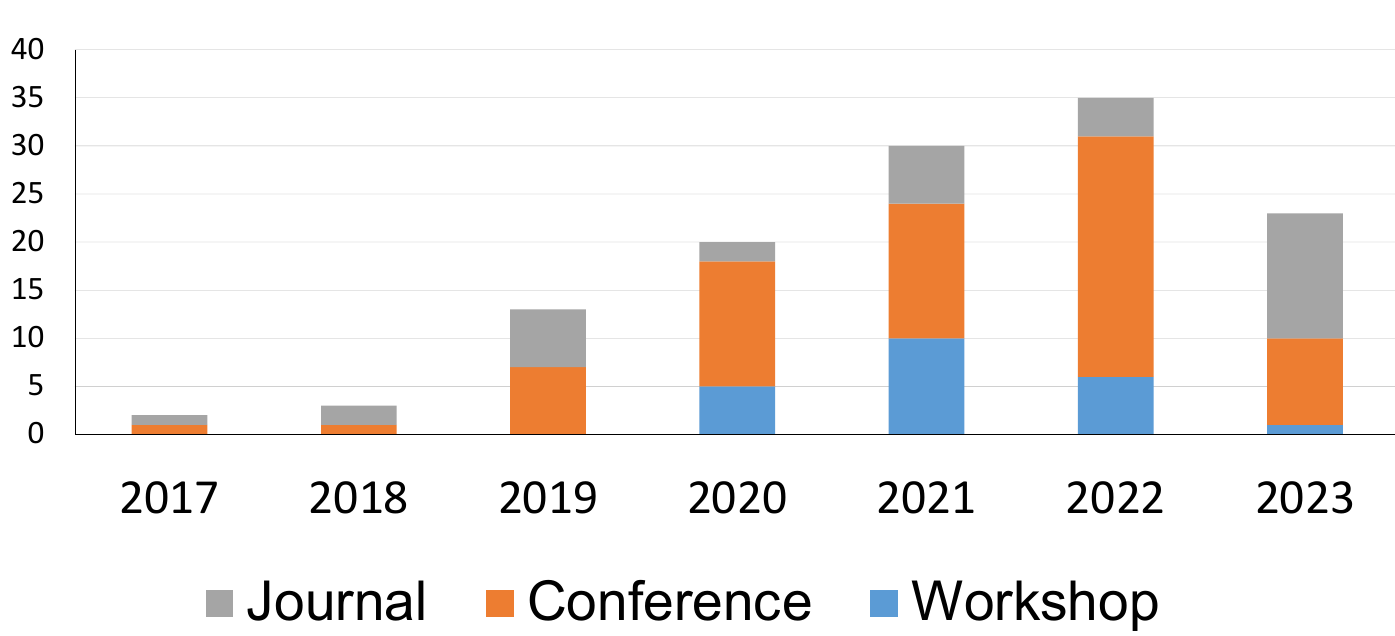}
 \caption{Yearly Publications Distribution}
 \label{fig:distribution}
\end{figure}

This RQ covers the publication trends over the years. Based on the earliest published primary study, we see this field emerged in $2017$. Although we started our search from the year $2010$, we found the first relevant paper in $2017$. As expected, there is a growing trend of studies in the field of RE for AI after that, as shown in Fig.~\ref{fig:distribution}.

\begin{figure*}
    \centering
    \includegraphics[width=\textwidth]{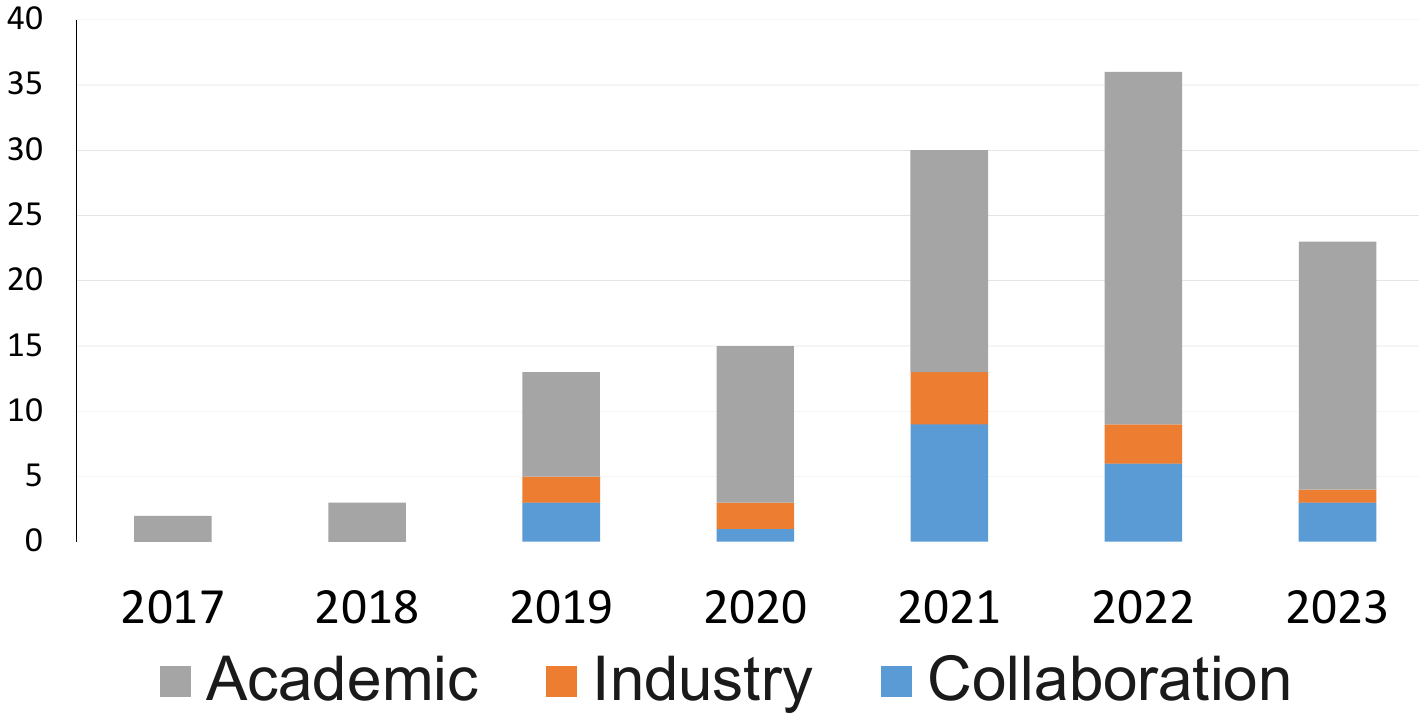}
    \caption{Research Community}
    \label{fig:Researchcommunity}
\end{figure*}

Initially, the exploration of this area was predominantly undertaken by the academic community. However, with the recent advancements in AI, there has been a noticeable increase in industrial engagement up to $2021$. This trend is evidenced by the rise in the number of publications from the industry during that period. Although there was a slight decrease in industrial publications in $2022$ and $2023$, the overall percentage of industrial papers has remained relatively steady since $2019$. This indicates a sustained interest and collaboration between academia and industry in RE4AI, as depicted in Fig.~\ref{fig:Researchcommunity}. Notably, IBM USA has emerged as a significant contributor with the highest publication count, while Fujitsu Laboratories Ltd., located in Kawasaki, Japan, has also made notable contributions to this field. 

The academic sector has consistently indicated substantial interest and research efforts in this field, with a gradual increase in publications over the years. In $2017$ and $2018$, the academic community published two and three papers, respectively. This momentum continued into $2019$ with eight academic articles, three collaborative efforts, and two industry contributions.

In $2020$, the output increased to $12$ academic publications, six collaborative articles, and two industry articles. This growth continued in $2021$, with $17$ academic articles, nine collaborative projects, and four industry publications. The upward trend continued in $2022$, reaching $27$ academic articles, 
six collaborative efforts, and two industry contributions.

By the end of July $2023$, early data indicates $19$ academic papers, three collaborations, and one industry publication. This sustained increase in academic involvement highlights the ongoing growth and interest in this research domain.

The increasing number of publications, especially from academia, shows growing interest and involvement in this research area. The steady yearly growth and numerous collaborations indicate an active and expanding research community. This ongoing momentum is evident even in the partial data for $2023$, highlighting the field's importance and the key role of the academic community in driving innovations forward.

Considering the publishing venue, $70$ out of $126$ ($54.9\%$) papers were published at various conferences, whereas, $28$ ($33.3\%$) were at workshops, and $28$ ($11.8\%$) were in journals. Around $2021$, workshops became more popular as an RE4AI venue, but conferences still account for a higher proportion. We can also observe that $23$ out of $28$ papers in the journals were contributed by the academic community. In contrast, the industry's preferred venues are conferences. 
The most recurring conferences in this domain are the International Requirements Engineering Conference (RE) with $7$ papers. It is followed by the International Working Conference on Requirements Engineering: Foundation for Software Quality (REFSQ)
, and the Conference on Human Factors in Computing Systems (CHI), with five papers each. Whereas in the workshop category, the most preferred venue is the International Requirements Engineering Conference Workshops (REW), followed by the Joint Proceedings of REFSQ-Workshops with nine and four publications, respectively.
Another main workshop is the Workshop on AI Engineering -- Software Engineering for AI (WAIN), which has two publications.
Moreover, only $28$ journal publications were found, out of which three were published in Requirements Engineering Journal and three in IEEE Computer.

In conclusion, we can see that the field of RE4AI has grown over the years, with a majority of papers published in conferences where workshops and journals have equal numbers. Furthermore, the industry has shown a significant interest and involvement in research within this field. Although less than half of the papers each year have industrial involvement, we believe that the industrial adoption of AI is experiencing consistent interest, primarily due to the significant adoption of Large Language Models (LLMs) within the industry. This trend is not only catalyzing a shift from conventional software paradigms towards AI-based applications but is also likely to amplify the demand for specialized RE approaches tailored to AI within the industrial sector. Consequently, traditional RE methodologies must be adapted to meet the unique demands and complexities of AI-based systems.

\subsection{Distribution of RE Topics for AI-based Systems (RQ2)}

Systematic Mapping Studies are typically employed to present a classification scheme for research topics within a specific field of interest. Analyzing the distribution of publications across these topics can provide insights into the breadth and depth of research, indicating the field's scope and its level of maturity.

 \begin{figure*}
    \centering
    \includegraphics[width=\textwidth]{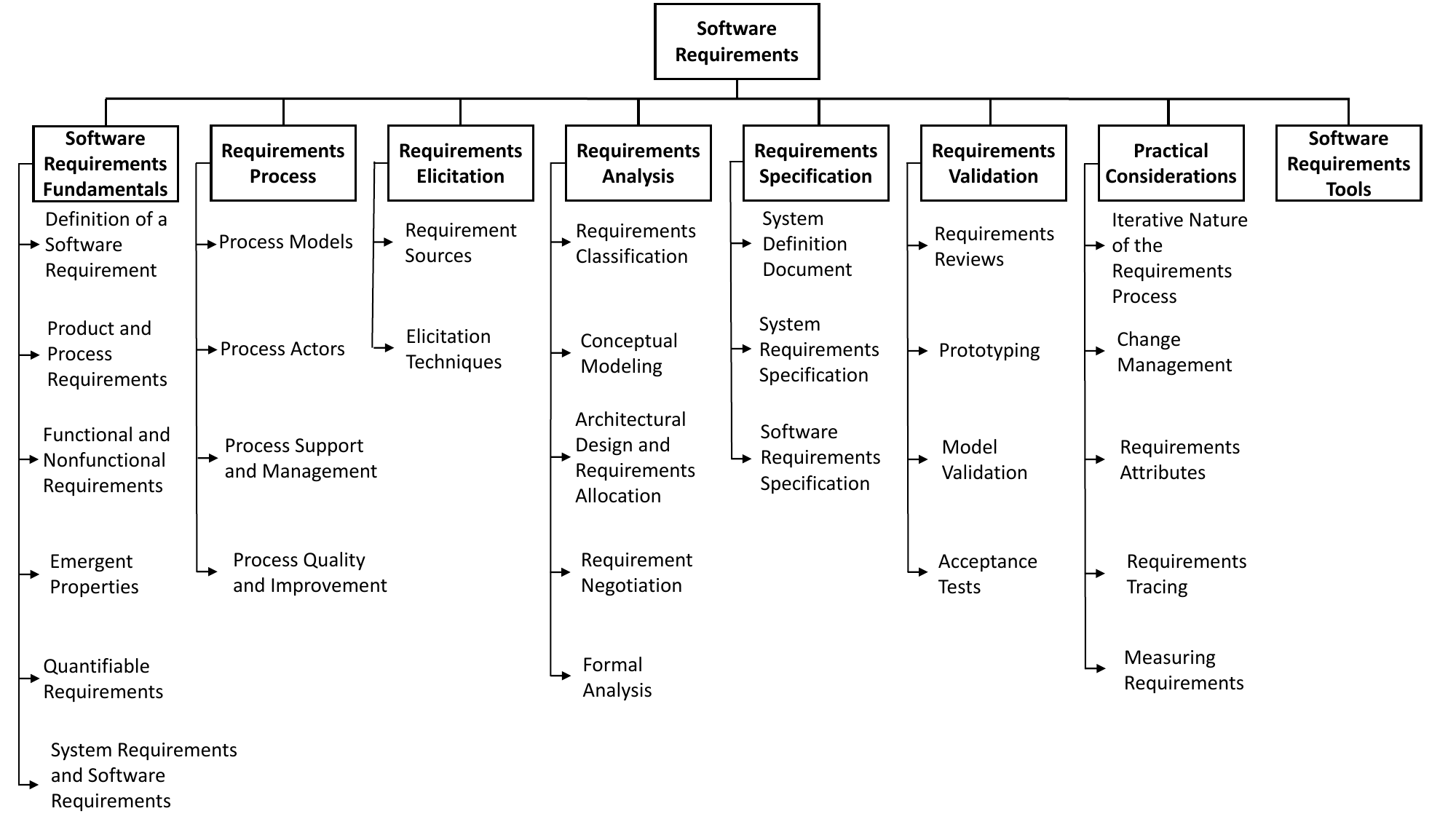}
    \caption{Breakdown of Topics for the Software Requirements KA \cite{bourque2014swebok}}
    \label{fig:swebok}
\end{figure*}
\begin{figure*}
    \centering
    \includegraphics[width=\textwidth]{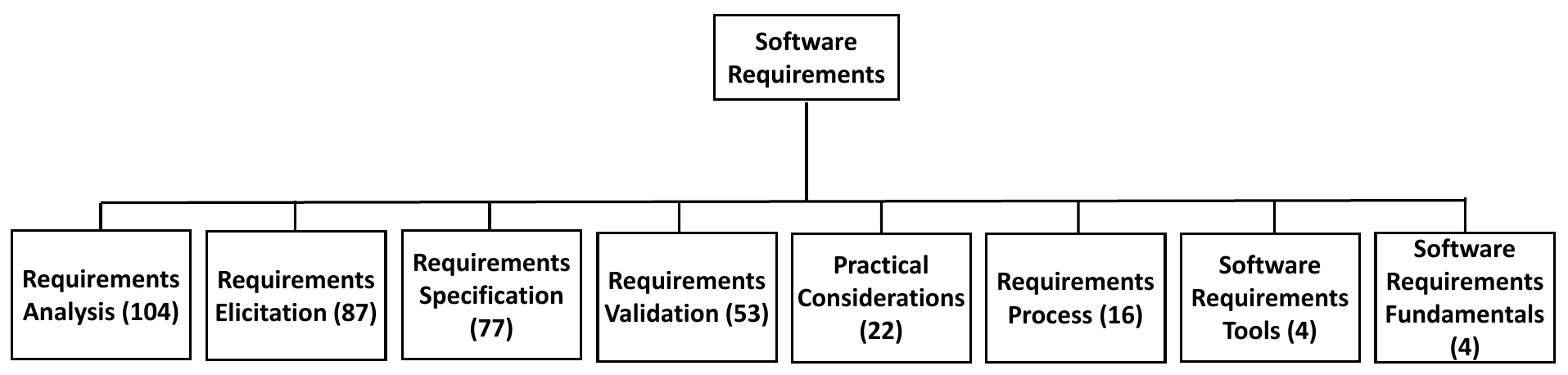}
    \caption{Number of papers in each category}
    \label{fig:RQ3-analysis}
\end{figure*}

To answer this research question, we used the well-established classification scheme of SWEBOK~\cite{bourque2014swebok} for RE topics as shown in Fig.~\ref{fig:swebok}. It allows us to observe where RE4AI research has been focused and which topics may still require attention. \textit{One paper can be classified under more than one topic, depending upon which RE topics they addressed in their research}. In addition, we analyzed topics and their sub-categories, such as which sub-topics have been addressed or remained unattended. It can help researchers identify further gaps in the current research landscape. Fig.~\ref{fig:RQ3-analysis} visualizes our general findings for this RQ.

Within our analysis, we identified that $104$ studies concentrated on requirements analysis, marking it as the predominant category in our classification. This category is notable for the introduction of $33$ RE practices, detailed in Section \ref{sec:newpractice}. The research work within this realm has primarily explored the integration of conceptual modeling \citeS{maass2021}, the classification of new(novel) requirement types \citeS{sheh2018}, and the assimilation of human-centric requirements into ML systems \citeS{shergadwala2021,mass2021}.

Moreover, our review reveals that $87$ studies were dedicated to requirements elicitation, representing the most substantial segment where established practices have been applied, as discussed in Section \ref{sec:existingpractices}. This highlights the field's ongoing efforts to refine and utilize traditional RE methodologies within the context of evolving technological frameworks. Furthermore, interviews \citeS{dhanorkar2021,brennen2020,weisz2021}, questionnaires \citeS{liao2020}, and scenarios have been used as practices in this area  \citeS{cirqueira2020,rivero2021,wolf2019}.

Further, we found that $77$ papers discussed topics related to requirements specification, with $30$ of these studies introducing new practices for specifying requirements. It is observed that the recent literature leans towards proposing practices specifically tailored to meet ML-related requirements, emphasizing stakeholder needs \citeS{suresh2021,gil2019towards}. Notably, only three studies adhered to existing practices for requirements specification.

Shifting the focus to the Requirements Validation Knowledge Area, $53$ studies were identified that delve into requirements validation, out of which eight introduced novel practices for conducting requirements validation. Thus, aiming to enhance the validation processes in line with AI-based systems.
Fig.~\ref{fig:RQ3-analysis} shows $22$ studies focused on \textit{practical consideration}, whereas $5$ studies proposed new practices for practical consideration. These studies highlighted requirements attributes regarding explanatory capabilities, ethical guidelines, and quality characteristics.
$16$ studies covered RE processes and focused on tailored RE processes for ML-based systems. These processes aim to incorporate ML-specific needs and additional types of requirements. Few studies highlighted the different perspectives in the business context during the RE process. We can observe that $13$ studies proposed new practices for \textit{RE process,} where $9$ applied existing RE processes to AI-based systems, primarily focusing on Goal-Oriented Requirements Engineering (GORE) \citeS{silva2019requirements,neace2018goal}.
The focus on \textit{requirements tools} has been relatively limited, with only $4$ studies identified in this domain proposing innovative tools to support the RE process for AI-based systems (see Section~\ref{sec:newpractice}). These tools are particularly aimed at streamlining tasks related to the elicitation and specification of requirements. 
In the last category, i.e., \textit{software requirements fundamentals}.  Only $4$ studies specifically addressed the topic of requirement definitions.

The trends we observed from these statistics suggest a field in transition, grappling with the unique challenges posed by AI and machine learning systems. The use of existing practices in elicitation and modeling points to a reliance on traditional RE strengths. However, the need for new practices, especially in analysis, specification, and the requirements process, suggests that AI-based systems introduce complexities and challenges that transcend the capabilities of traditional RE methods. These new practices likely address AI-specific concerns such as ethical considerations, data quality and sourcing, model transparency, and the dynamic nature of learning algorithms.

\subsection{RE4AI Research Maturity (RQ3)}

To address this question, we classified the papers according to the taxonomy by Wieringa et al.~\cite{wieringa2006requirements}. 

We aim to highlight the research methods so far used by researchers in the RE4AI directions and how these practices have been evaluated.

 \begin{figure}
    \centering
    \includegraphics[width=\textwidth]{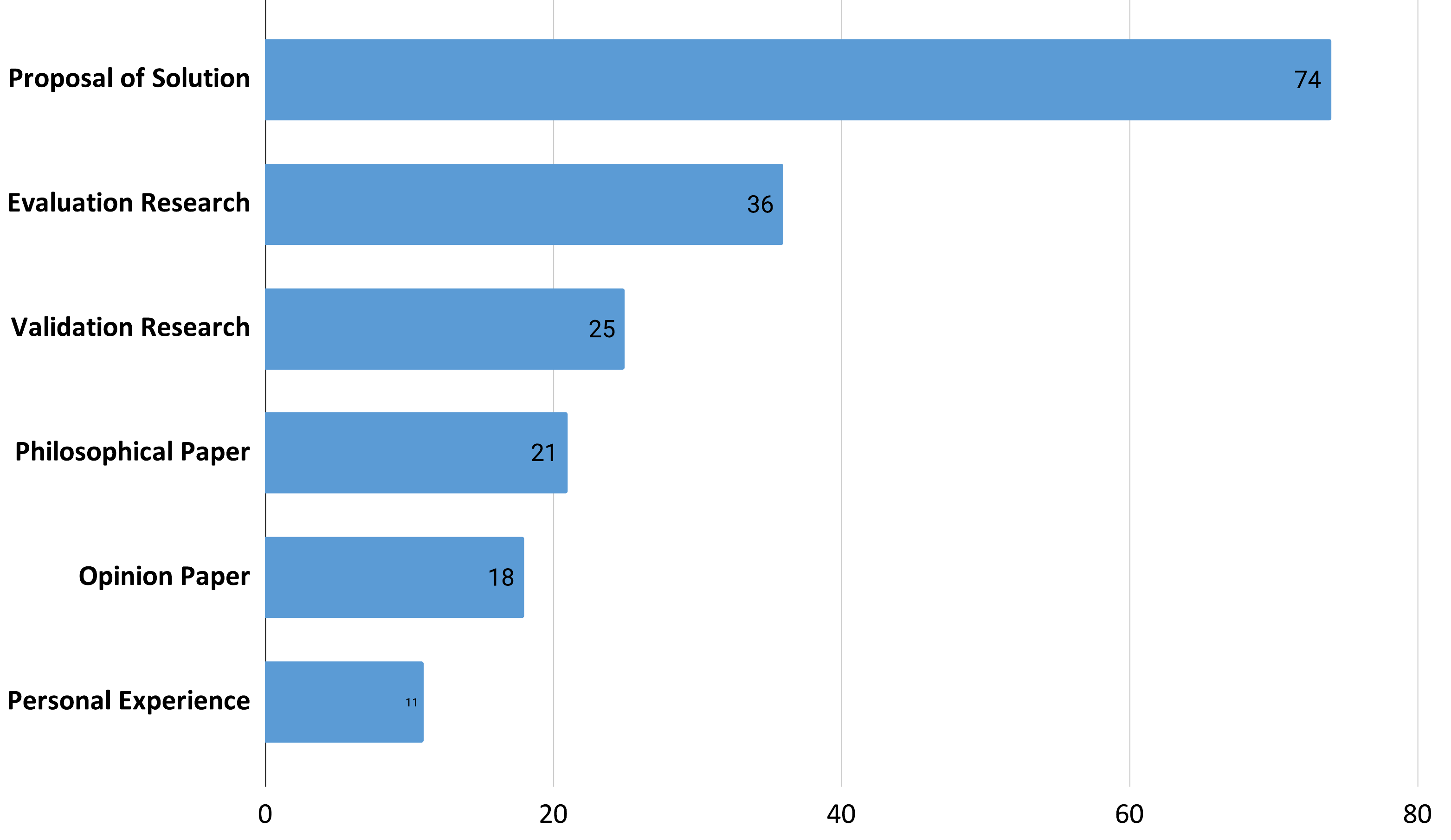}
    \caption{Distribution of papers according to Wieringa's classification}
     \label{RQ4}
\end{figure}

Fig.~\ref{RQ4} shows that $74$ studies fall into the \textit{proposal of solution} category, i.e., papers proposing a solution and establishing its relevance. Either the proposed solutions should be novel, or an existing solution should be adapted and applied to a new domain.
A new conceptual framework has been proposed in $21$ studies, and we classify them as \textit{philosophical papers}, as some of them do not provide a direct solution but all of them offer a new way of understanding and categorizing requirements.

Papers that investigate proposed solutions' properties while the solution still requires implementation in RE are classified as \textit{validation research}. $25$ papers are in this category that validate a solution proposed in the same paper or elsewhere.
Papers that apply RE techniques in practice or investigate the usage of RE practices are classified as \textit{evaluation research}. The novelty of the practice is not essential in this case. Instead, the knowledge claim of the paper should be novel. $18$ among our primary studies describe the authors' position, primarily to provoke discussions about RE4AI topics. These types of papers are categorized as \textit{opinion papers}. Lastly, $11$ studies reported personal experiences and were labeled as \textit{personal experience}. Papers could be classified with more than one of these categories.

 \begin{figure*}
    \centering
    \includegraphics[width=5in]{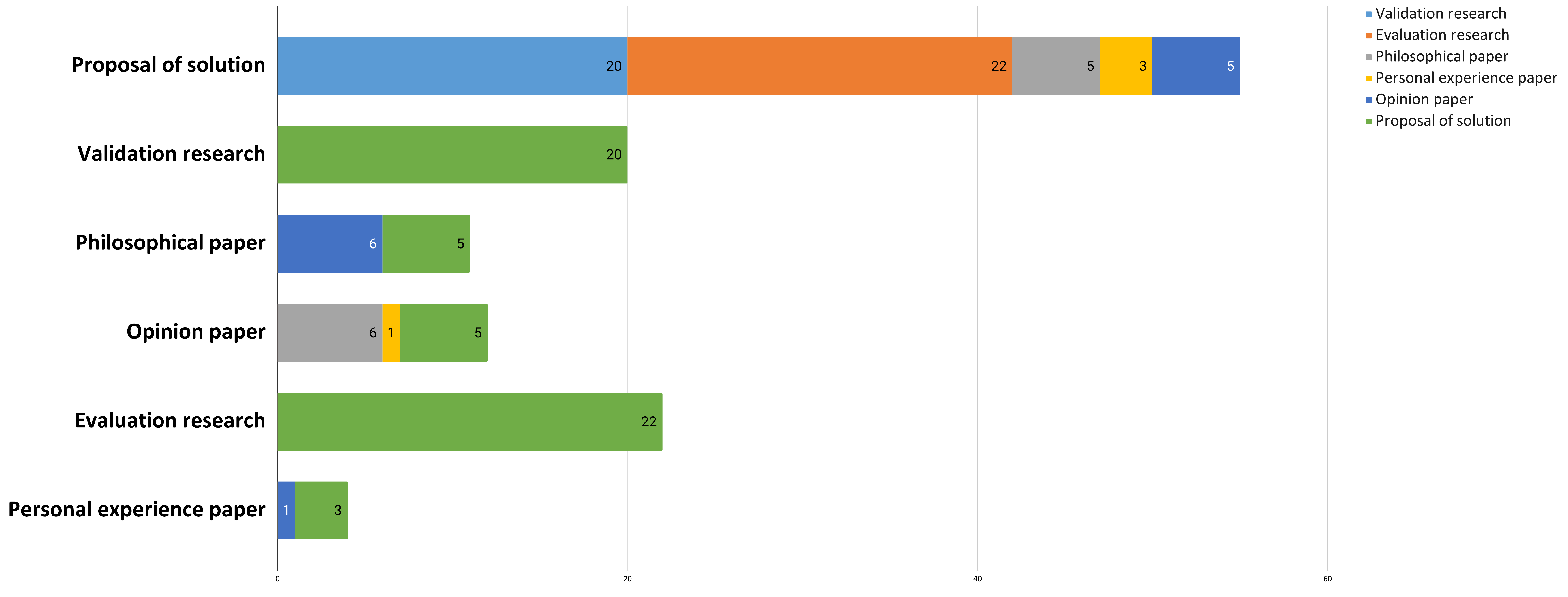}
    \caption{Frequently Combined Research Methods}
    \label{fig:research-facet}
\end{figure*}
Our analysis (See Fig.~\ref{fig:research-facet}) reveals a notable trend in research practices. Specifically, we observe that \textit{proposal of solution} papers predominantly incorporate validation research, with 20 studies validating solutions and 22 engaging in evaluation research. Additionally, 5 papers combine proposals of solutions with opinion and philosophical discourse, while 3 include personal experiences. In \textit{validation research}, a common pattern emerges where 20 studies both propose and validate solutions, highlighting a preference for self-validation. Philosophical and opinion papers often intertwine, sharing a focus on conceptual framework. Evaluation methods vary, with case studies ($43$) being predominant, followed by surveys ($15$) and minimal use of mixed methods ($1$). This reflects a broader inclination towards practical validation in the proposal of a solution, while opinion and experience papers typically lack such research.

\begin{figure*}
    \centering
    \includegraphics[width=\textwidth]{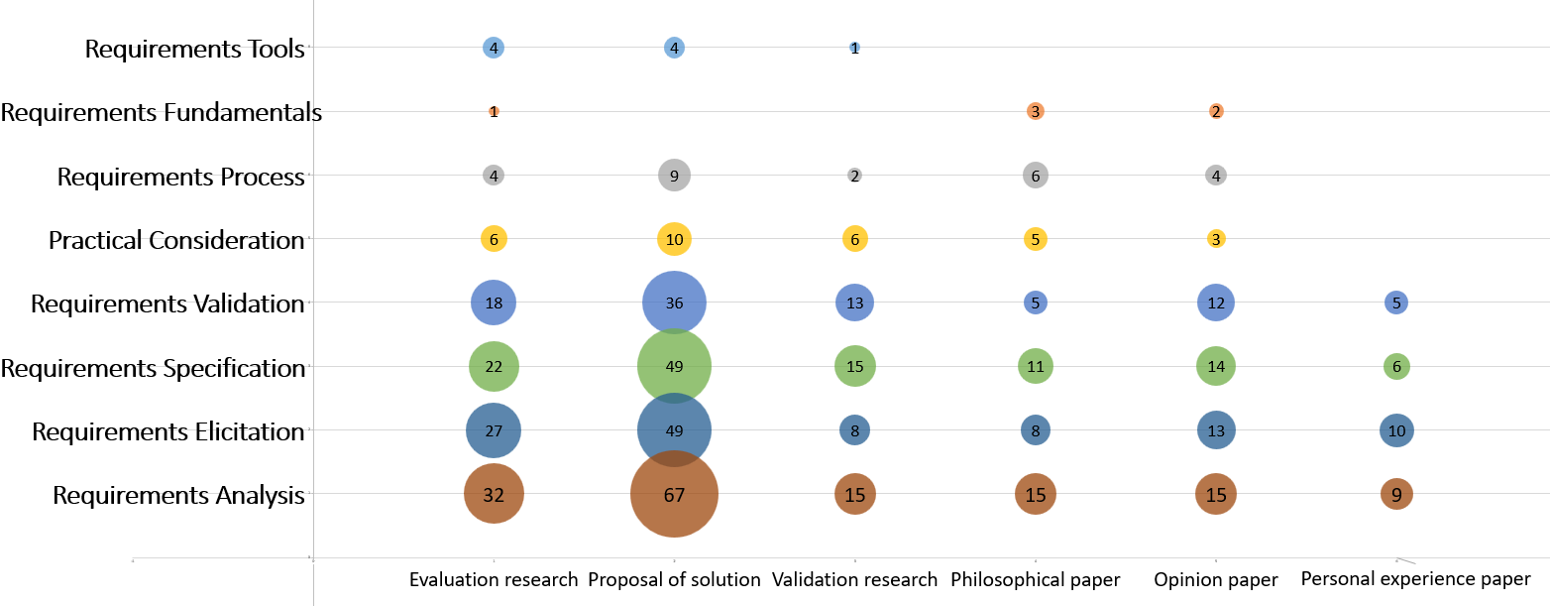}
    \caption{Maturity of the research area by analyzing the prevalent research methods utilized in each KA}
    \label{fig:RQ3-4}
\end{figure*}

Further, to assess the maturity of the research, we investigate which type of research is conducted in each SWEBOK KA.
Fig.~\ref{fig:RQ3-4} shows how RE SWEBOK~\cite{bourque2014swebok} topic are addressed using Wieringa's classification~\cite{wieringa2006requirements}. It should be noted that one paper can be in more than one publication type, so the total adds up to more than 126.
This analysis highlights that significant focus has been placed on requirement analysis, elicitation, specification, and validation. However, foundational aspects of requirements, such as their fundamental principles, processes, and practical considerations, have been overlooked. Additionally, there is a notable shortage of tools to support the development of AI-based systems. The data reveals that while the initial activities of RE receive considerable attention, there is still a deficiency in managing the overall RE process for AI-based systems effectively

\subsection{RE Practices for AI-based Systems (RQ4)}

This question aims to highlight the use of existing RE practices and the direction in which new RE practices specific to AI have emerged. We extracted the practices according to the classification scheme provided in Section~\ref{sec:RQ}.

\begin{figure}
    \centering
    \includegraphics[width=2.6in]{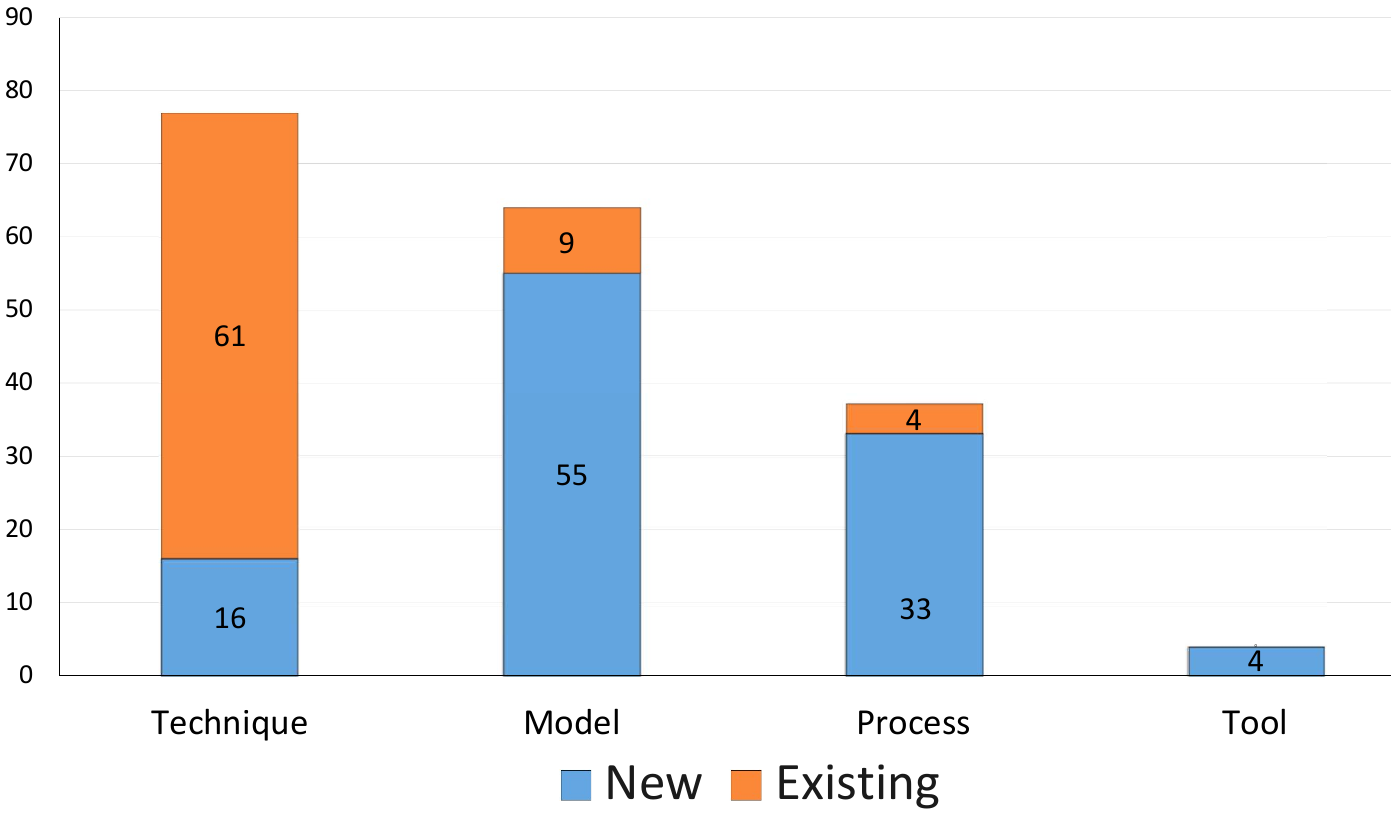}
    \caption{Existing vs new practices}
    \label{fig:new-vs-exisiting}
\end{figure}

The bar chart in Fig.~\ref{fig:new-vs-exisiting} indicates literature proposing new processes, techniques, models, and tools. Though more research is focused on techniques, the novelty can be seen more in model, process, and tool-related research. It is evident that current techniques are more frequently used.
One major takeaway is that existing RE techniques could be adapted for AI-based systems, however, standard RE processes and tools are not adapted for the development of AI-based systems. In the subsequent sections, we will elaborate on how existing practices have been used and what new practices have been used.

\subsubsection{Usage of Conventional RE Practices for AI}
\label{sec:existingpractices}
We analyzed the suitability of current RE practices for AI by determining what RE practices have been used for such systems. 
The resulting model is shown in Table~\ref{Existing practices RQ2}.
We group the practices according to the RE topics in SWEBOK~\cite{bourque2014swebok}. Since we found \textit{requirements modeling}, which is not part of SWEBOK, to be an important topic among our papers, we included it as a distinct group. 
Further, we found \textit{requirements elicitation}, \textit{requirements process}, \textit{requirements validation}, \textit{requirements analysis}, and \textit{requirements specification} KAs using existing practices of RE. Each paper may have multiple RE practices and could fall into multiple software requirements KAs.

\begin{table*}
\centering
\caption{Existing Practices Used}
\label{Existing practices RQ2}
\resizebox{\columnwidth}{!}{%
\begin{tabular}{llll}
\hline
\textbf{RE topic} & \textbf{Practices used} & \textbf{Type} & \textbf{Paper ID} \\ \hline
\multirow{14}{*}{Requirements Elicitation (40)} & Interviews & Technique & \begin{tabular}[c]{@{}l@{}}\citeS{dhanorkar2021}, \citeS{brennen2020}, \citeS{weisz2021}, \\ \citeS{rincon2021speaking}, \citeS{shergadwala2021}, \citeS{rivero2021}, \\ \citeS{konigstorfer2021software}, \citeS{habibullah2023non}, \citeS{weinert2023exploring} \\ \citeS{yu2022stakeholders}, \citeS{gruning2023towards}, \citeS{gabriel2022requirements} \\
\citeS{maalej2023tailoring}, \citeS{elshan2022requirements}, \citeS{balasubramaniam2023transparency} \\ \citeS{nahar2022collaboration} \end{tabular} \\
 & Scenario-based requirements elicitation & Technique & \begin{tabular}[c]{@{}l@{}}\citeS{cirqueira2020}, \citeS{rivero2021}, \citeS{wolf2019},\\  \citeS{jansen2021}, \citeS{heck2023defining}, \citeS{cirqueira2020scenario} \\
 \citeS{gauerhof2020assuring}
 \end{tabular} \\
 & Card sorting & Technique & \citeS{drozdal2020}, \citeS{siqueira2022guide} \\
 & Controlled experiment & Technique & \citeS{drozdal2020} \\
 & Focused groups & Technique & \begin{tabular}[c]{@{}l@{}}\citeS{brennen2020}, \citeS{maalej2023tailoring}, \citeS{naveed2022explainable} \\
 \citeS{cerqueira2021exploring}
 \end{tabular}\\
 & SigniFYIng message & Model & \citeS{jansen2021} \\
 & Survey & Technique & \begin{tabular}[c]{@{}l@{}}\citeS{de2022requirements}, \citeS{yu2022stakeholders}, \citeS{gabriel2022requirements}\\ \citeS{eichelberger2022requirements}, \citeS{elmesalawy2021ai}, \citeS{naveed2022explainable}\end{tabular} \\
 & Think aloud & Technique & \citeS{drozdal2020} \\
 & User Stories & Technique & \citeS{siqueira2022guide}, \citeS{maalej2023tailoring} \\
 & Questionnaire & Technique & \citeS{liao2020}, \citeS{weinert2023exploring} \\ \hline
\multirow{13}{*}{Requirements Modelling (26)} & Conceptual modelling & Technique & \begin{tabular}[c]{@{}l@{}}\citeS{lukyanenko2019}, \citeS{mass2021}, \citeS{maass2021}, \\ \citeS{nalchigar2018}, \citeS{nalchigar2021}, \citeS{heck2023defining} \\
\citeS{shao2022data}, \citeS{maass2022conceptsuperimposition}
\end{tabular} \\
 & \begin{tabular}[c]{@{}l@{}}AMDiRE model \\ (Artefact Model for Domain-independent RE)\end{tabular} & Model & \citeS{chuprina2021} \\
 & Data flow diagrams & Technique & \citeS{wilhjelm2020} \\
 & ISO 26262 & Model & \citeS{salay2019} \\
 & Persona & Technique & \citeS{rivero2021}, \citeS{noda2021requirement} \\
 & \begin{tabular}[c]{@{}l@{}} actor-based requirements Modelling using \\ istar framework \end{tabular} & Technique & \citeS{barclay2021identifying}, \citeS{barrera2022use}, \citeS{barrera2021applying} \\
 & Goal Modeling & Technique & \citeS{li2023dealing}, \citeS{husen2022traceable} \\
 & STRIDE & Technique & \citeS{wilhjelm2020} \\
 & Model Driven Engineering & Model & \citeS{jahic2023semkis}, \citeS{bergelin2022industrial} \\
 & UML & Technique & \begin{tabular}[c]{@{}l@{}} \citeS{ries2021}, \citeS{dalpiaz2020}, \citeS{husen2022traceable} \\ \citeS{silva2019requirements} \end{tabular} \\ \hline
\multirow{2}{*}{Requirements Process (9)} & GORE & Technique & \begin{tabular}[c]{@{}l@{}} \citeS{belani2019}, \citeS{ishikawa2020evidence}, \citeS{yu2021}, \\ \citeS{tun2021goal}, \citeS{khan2022handling} , \citeS{kaindl2020towards} \\
\citeS{neace2018goal}, \citeS{silva2019requirements}
\end{tabular} \\
 & Softgoal Interdependency Graphs (SIGs) & Technique & \citeS{kohl2019} \\ \hline
\multirow{3}{*}{Requirements validation (3)} & Document analysis & Technique & \citeS{rivero2021} \\
& Fuzzy Kano & Technique & \citeS{chang2023stakeholder} \\
 & ISO 25012 model & Model & \citeS{challa2020faulty} \\ \hline
\multirow{3}{*}{Requirements Analysis (3)} & ISO 25000 series, known as SQuaRE & Model & \citeS{kuwajima2019} \\ 
& ISO 26262 and ISO/FDIS 21448 & Process & \citeS{acar2022application} \\
& Use case boundary condition & Technique & \citeS{maalej2023tailoring} \\ \hline
\multirow{3}{*}{Requirements Specification (3)} & Alloy formal specification & Technique & \citeS{ries2021} \\
& EARS & Technique & \citeS{heck2023defining} \\ 
& Operational Design Domain & Technique & \citeS{habibullah2023requirements}
\\ \hline
\end{tabular}%
}
\end{table*}

\textbf{Requirements Elicitation:} 
Out of $126$ papers examined, $40$ employed various practices for eliciting requirements. 
Among these practices, interviews emerged as the predominant method, with $16$ of the $40$ papers utilizing them for gathering requirements. 
Notably, a significant number of researchers favored semi-structured interviews as a tool to initiate conversations around generative AI~\citeS{weisz2021} and to foster a collaborative design process \citeS{rincon2021speaking,shergadwala2021}. Similarly, researchers \citeS{rincon2021speaking,rivero2021,dhanorkar2021,brennen2020} also conducted semi-structured interviews for requirements elicitation. Additionally, there are studies~\citeS{gruning2023towards,gabriel2022requirements,maalej2023tailoring,nahar2022collaboration} concentrated on requirements elicitation that prioritize stakeholders' perspectives and needs.

Other methods identified for requirements elicitation across the reviewed literature encompass surveys, scenario-based elicitation techniques~\citeS{rivero2021, wolf2019, jansen2021, heck2023defining, cirqueira2020scenario, gauerhof2020assuring}, questionnaires~\citeS{liao2020, weinert2023exploring}, think-aloud protocols~\citeS{drozdal2020}, focus groups~\citeS{brennen2020, maalej2023tailoring, naveed2022explainable, cerqueira2021exploring}, and controlled experiments, showcasing a diverse range of strategies for gathering and understanding project requirement.

\textbf{Requirements Modelling:} Twenty-six among $126$ papers mentioned an existing practice for requirements modeling. The most frequently used practice for this topic was \textit{conceptual modeling,} which $8$ different articles have addressed. In \citeS{nalchigar2021}, conceptual modeling is used for requirements elicitation, design, and development of ML solutions. Similarly, \citeS{maass2021} incorporated conceptual modeling into a data science project and applied it to a healthcare application. The authors of \citeS{nalchigar2018} use conceptual modeling to illustrate the business view, analytics design view, and data preparation view. These perspectives are used to relate the corporate strategy to analytics algorithms and data preparation operations. Authors in \citeS{lukyanenko2019} argued conceptual modeling could support the application of ML within an organization while improving usability and optimizing the performance of ML algorithms. Additionally, in \citeS{mass2021} the authors demonstrated that conceptual modeling can be used to map human mental models to model AI-based systems. Other frequently used modeling techniques are \textit{scenario-based design} and the \textit{Unified Modeling Language (UML)}. Husen et al.~ \citeS{husen2022traceable} use UML for analyzing ML safety requirements top-down from higher-level
business requirements, whereas~\citeS{silva2019requirements} provides comparison using UML diagrams and aims to propose effective design practices for planning problems, with a focus on the early requirements phase.

Furthermore, within the scope of the \textbf{requirements process}, Goal-Oriented Requirements Engineering (GORE) and Softgoal Interdependency Graphs (SIGs) have been employed in $8$ and $1$ studies, respectively, as detailed in Table~\ref{Existing practices RQ2}.
 It's also worth noting that there has been a lesser focus on utilizing existing practices for \textbf{requirements validation, specification}, and \textbf{analysis}, with only $3$ instances identified in each of these areas. It can also be observed that the researchers paid attention to using existing ISO standards, including ISO 26262 \citeS{acar2022application}, ISO 25012 \citeS{challa2020faulty}, and ISO 25000 \citeS{kuwajima2019}.

\subsubsection{New Practices Employed in RE4AI}
\label{sec:newpractice}
For new practices, Table~\ref{tab:RQ2New} provides a brief description of each practice and the type of practice, i.e., model, process, technique, or tool. We can observe that numerous studies have proposed new models, while a significant number also introduced new processes.
Further, we will elaborate on how each KA has been addressed by researchers.



\tiny
\begin{longtable}{>{\arraybackslash}p{1.6cm}>{\arraybackslash}p{0.5cm}>{\arraybackslash}p{1cm}>{\arraybackslash}p{6.6cm}}
\caption{New Practices Proposed}
\label{tab:RQ2New}\\

\toprule

 \centering \textbf{Contribution in  SWEBOK KA} &  \textbf{Paper ID} &  \textbf{Type} & 
 {\centering \textbf{Short Description}}  \\ 

\midrule

Requirements Analysis (33) & \citeS{belani2019}&	Model&	RE4AI taxonomy for building AI-based complex systems \\
&\citeS{thinyane2020} &	Technique &	Multi-aspectual analysis of AI Ethics frameworks \\
&	\citeS{sheh2018}	 & Model &	Novel categorisation for explanations\\
& \citeS{shergadwala2021}	 & Model &	Investigating human-centric design requirements in engineering design \\
& \citeS{maass2021}	 &Process &	A framework for incorporating conceptual Modeling into data science projects \\
&\citeS{yu2021}&	Process&	Propose an RE framework needs to address both sides of the cognitive cycle for all relevant actors. \\
&\citeS{hall2019} &	Process &	Presents a five-step systematic method in the development of an explainable AI (XAI) system \\
&\citeS{kohl2019}&	Model	& Conceptual analysis which unifies the different notions of explainability and the corresponding explainability demands \\ 
&\citeS{suresh2021}&	Model	&Framework with a more granular and composable vocabulary to characterize the stakeholders of interpretable ML\\ 
&\citeS{wang2019}&	Model& Framework for XAI researchers and designers to identify pathways along which human cognitive patterns drive needs for building XAI and how XAI can mitigate common cognitive biases.\\

&\citeS{camilli}&	Process&	Risk Modeling approach tailored to Collaborative AI systems\\
&\citeS{mass2021}&	Process&	Propose a framework for progressing from human mental Models to machine learning Models and implementation via the use of conceptual Models \\
&\citeS{gil2019towards}&	Model&	Task analysis of human-guided machine learning \\
&\citeS{barclay2021identifying}& 	Model& A framework and an actor-based requirements model to analyze the roles, requirements, and responsibilities in AI ecosystems, enhancing stakeholder identification and revealing potential goal tensions.\\
&\citeS{battistoni2023can}&	Model&	Presents a list of AI requirements for effective human-AI interaction \\
&\citeS{weinert2023exploring}&	Model	& Development of a requirements model that captures diverse and sometimes conflicting needs of patients and AI researchers/developers for the pAItient project, facilitating communication of stakeholder requirements within the consortium. \\

&\citeS{lavalle2022law}&	Model&	Introduces a dynamic framework that integrates project context and law modeling to identify protected attributes for AI model fairness, maps legal requirements to dataset attributes, aids in selecting suitable fairness definitions, and visually represents AI model outputs for fair and accurate decision interpretation. \\
&\citeS{habibullah2023exploring} &	Model&	Identify challenges related to NFRs and develop solutions to manage NFRs for ML systems \\
&\citeS{liubchenko2022requirements}	&Technique&	Method for evaluating risks in AI/ML-based software systems design, focusing on dependency-driven assessments \\
&\citeS{elmesalawy2021ai}	&Model&	Summarizes the refined Lifelong Learning System (LLS) requirements based on feedback from students and instructors, \\ 
&\citeS{tun2021goal}	&Model&	Introduces a goal-centralized meta-model for integrating FRs and NFRs through goal-oriented analysis of ML systems \\
&\citeS{nitta2022ai}	&Process&	Identifies AI's ethical problems by linking responsible AI requirements from ethics guidelines with AI system interactions, including building an AI ethics model to embody guidelines as requirements \\
&\citeS{shao2022modeling}	&Technique&	 Presents a two-tiered approach for data requirements modeling in ML systems within the scope of requirements analysis. The foundational tier maps out the learning context, employing a feature-oriented domain analysis to detail system components, their environment, and interrelations. The subsequent tier advances to define property-based specifications.\\
&\citeS{lee2023provenance}	&Process&	Introduces a provenance-based, trust-aware RE framework for self-adaptive systems, enabling engineers to derive trust-aware goal models from user requirements \\
&\citeS{heyn2023investigation}	&Model&	Outlines challenges in specifying training data and runtime monitoring for ML models faced by practitioners, identifies interconnections between these challenges, and offers recommendations to address the root causes \\
&\citeS{chang2023stakeholder}	&Process&	Proposes a systematic approach for evaluating stakeholder requirements  \\
&\citeS{husen2022traceable}	&Process&	Outlines a top-down approach for analyzing machine learning safety requirements \\
&\citeS{guizzardi2020ethical}&	Technique &		 Leveraging established RE techniques for software systems to elicit and analyze ethical requirements \\
&\citeS{kaindl2020towards}&	Model&	This paper proposes a new theoretical framework for an extended requirements problem, incorporating both stakeholder goals and objectives specific to the AI-based system \\
&\citeS{naveed2022explainable}	&Model&	Expands current explanation categories to suit the financial domain by pinpointing specific explainability needs of users \\
&\citeS{maass2022conceptsuperimposition}&	Model&	Introduces a framework for transitioning from human mental models to machine learning models through conceptual models \\
&\citeS{barrera2021applying}&	Model	&Introduce an extension of i*, addressing the disconnect between machine learning and conceptual modeling to establish a foundational methodology for machine learning requirements engineering \\
&\citeS{li2023dealing}&	Model& 	Presents an explainability framework that automates the recommendation of explainability methods for system design, enhancing system explainability and efficiency for developers, and mitigating the tension between explainability and usability, \\
\midrule

Requirements Process (13) 	&\citeS{ishikawa2020evidence}	& Process& 	Evidence-driven RE to deal with the additional type of uncertainty of the implementation \\
	&\citeS{vogelsang2019}&	Process	&Tailored RE methodology for ML systems incorporating New types of requirements as well \\
	&\citeS{ries2021}&	Process&	Model-driven engineering method based on traditional RE \\
	&\citeS{camilli2021}&	Process	&RE to  build effective ML systems with provable compliance assurances \\
	&\citeS{nalchigar2018}&	Process&	Modelling framework for requirements analysis and design from a business view, analytics design view, and data preparation view \\
	&\citeS{chuprina2021}&	Model&	An artefact-based RE approach for the development of datacentric systems \\
	&\citeS{qadadeh2020}&	Process	&Developed an improved Agile data mining framework to fulfill the government business objectives and needs and a  systematic way for identifying business problems \\
	&\citeS{treacy2022legal}&	Process&	Methodology for developing and assessing legal, privacy, social, and ethical requirements \\
	&\citeS{franch2023requirements}& Process	&Requirements Engineering for AI \\
	&\citeS{habiba2022can}&	Process&	Outlines challenges in XAI and proposes a framework with research directions for using RE practices to address these challenges. \\
	&\citeS{zhang2021dde}&	Process&	Introduces the Data-Driven Engineering process, a systematic approach for ML application in industry, featuring hierarchical RE and semi-automated data set generation, integrating with other processes in a V-Model, aiming at harmonizing development levels and automating dataset compilation \\
	&\citeS{gauerhof2020assuring}&	Process&	Safety requirements can be systematically and traceably generated and refined across the different life-cycle phases of the MLM, \\
	&\citeS{neace2018goal}&	Process	& Introduces a GORE-based methodology for autonomy requirements engineering in Unmanned Aircraft Systems. \\
\midrule

Requirements Specification (30) 	& \citeS{kuwajima2019}&	Model&	Specification of safety requirements based on uncertainty\\
	&\citeS{hu2020}	&Process	&Approach for specifying and testing requirements\\
	&\citeS{singh2021}	&Technique	& Efficiently tackle the issue and derive component-level requirements \\
	&\citeS{suresh2021}	&Model&	Framework with a more granular and composable vocabulary to formulate the stakeholders' needs\\
	&\citeS{salay2019}	&Process&	Requirements for specification languages and identify types of specification that are well-suited to ML-based components\\
	&\citeS{rahimi2019}	&Process&	Proposed an approach to improve the process of requirements specification in which an MLC is developed and operates by explicitly specifying domain-related concepts\\
	&\citeS{gil2019towards}	&Model&	Human-guided machine learning as a hybrid approach where a user interacts with an AutoML system and tasks it to explore different problem settings that reflect the user’s knowledge about the data available\\
	&\citeS{schuh2022case}	&Model&	Synthesizes 75 unique requirements from a case study\\
	&\citeS{konigstorfer2021software}	&Model&	Identified five requirements for AI documentation that emphasize the need for engineers to combine technical details with understandable integration into business processes.\\
	&\citeS{battistoni2023can}	&Technique&	Introduces Intelligence-Centered Design (ICD), a modification of the Human-Centered Design approach, to integrate AI considerations from the start, providing guidance for novice designers in creating AI-based interactive systems with a focus on AI-human interaction and user experience.\\
	&\citeS{habibullah2023non}	&Technique&	Specifying NFR for the whole system\\
	&\citeS{berry2022requirements}	&Model&	Emphasizes that requirements specification for AI systems must include not only evaluation measures (M1, ..., Mn) but also criteria for acceptable values, their relative importance for trade-offs, vetting processes, and essential data like training data for proper AI functionality\\
	&\citeS{fernandez2023trustworthy}	&Model&	Introduced a comprehensive framework for AI trustworthiness, emphasizing a human-centric approach with criteria like human agency, security, privacy, and fairness.\\
	&\citeS{gruning2023towards}	&Model&	Holistic view of the specific requirements for AI-enabled medical devices\\
	&\citeS{bartlett2022characterizing}	&Technique&	Define sensor accuracy requirements for medical devices using ML algorithms for stability scoring\\
	&\citeS{wang2022framework}	&Model&	Paper proposes a specification framework for ML requirements\\
	&\citeS{maalej2023tailoring}	&Technique&	Proposed Quality levels for requirements specifications\\
	&\citeS{elshan2022requirements}	&Model&	Propose a set of requirements for an AI-based team member\\
	&\citeS{alagarswamy2023towards}	&Model&	List of generic audit requirements, which are technically relevant to assure the trustworthiness, security, safety, robustness, and explainability,\\
	&\citeS{vaida2021user}	&Model&	List of shared Requirements\\
	&\citeS{eichelberger2022requirements}	&Model&	67 usage view activities/scenarios, 141 top-level requirements, and 179 detailed sub-requirements\\
	&\citeS{noda2021requirement}	&Model&	Constructs user personas for AI medical interviews, from which it derives specific usability, reliability, and acceptability requirements, employing ISO/IEC 25010:2011 standards to address previously overlooked aspects\\
	&\citeS{jovanovic2022explainability}	&Technique&	Specifying  and formulating user requirements for explainable AI\\
	&\citeS{villamizar2022towards}	&Process&	Proposes a perspective-based method for specifying ML-enabled systems\\
	&\citeS{jahic2023semkis}	&Model&	Introduces SEMKIS-DSL, a textual domain-specific language designed to assist software engineers in specifying the requirements\\
	&\citeS{pinto2021requirement}	&Technique&	Provides a list of promising techniques for requirement specification, validation, and verification.\\
	&\citeS{balasubramaniam2023transparency}	&Model	&Introduces a model and template for defining explainability requirements in AI systems\\
	&\citeS{bergelin2022industrial}	&Model&	Definition of 78 high-level requirements refined into 30 generic ones by AIDOaRt partners for advancing cyber-physical systems development using AI, DevOps, and Model-driven engineering \\
 
	&\citeS{kemell2022utilizing}	& Technique&	Introduces Ethical User Stories (EUS) as a method to integrate AI ethics into standard SE practices, enabling the formulation of both FRs and NFRs\\
	&\citeS{bell2023think}	&Model&	Proposes a transparency playbook for technologists, aimed at developing AI systems that adhere to legal and regulatory standards and satisfy user requirements\\
\midrule
 Requirements Elicitation (16) 	&\citeS{nakamichi2020requirements}&	Model	&A method to identify requirements \\
	&\citeS{wolf2019}	&Model&	Scenario-based design for XAI: "Explainability scenarios focus on what people might need to understand about AI systems\\
	&\citeS{liao2020}	&Model	&Present an extended XAI question bank by combining algorithm-informed questions and user questions\\
	&\citeS{jansen2021}	&Model&	Scenario based explainability\\
	&\citeS{chari2020}	&Model	&Ontology to support user requirements for explanations in the domain of healthcare\\
	&\citeS{nguyen2021holistic}&	Process&	Methodology for addressing explainability requirements in ML services for IoT Cloud systems through stakeholder involvement, end-to-end engineering processes, and multiple aspects of explainability.\\
	&\citeS{ahmad2023requirements}&	Model&	Provide a catalog to elicit requirements and a conceptual Model to present them visually.\\
	&\citeS{siqueira2022guide}	&Process&	Guide for Artiﬁcial Intelligence Ethical Requirements Elicitation\\
	&\citeS{gabriel2022requirements}	&Technique&	Technique for a socio-technical requirements elicitation in the design of AI-based systems by adapting the HTO-analysis.\\
	&\citeS{dey2023multi}	&Process	&Introduces a multi-layered framework with a verifiable template for eliciting data requirements and uses Dempster-Shafer theory to assess training data quality through expert judgments,\\
	&\citeS{al2022resam}	&Process	&A requirements process that amalgamates insights from domain experts, forums, and formal documentation to identify and articulate requirements and design definitions as time-series attributes, enhancing the development of deep learning-based anomaly detectors\\
	&\citeS{silva2022requirements}	&Model	&Explore the requirements elicitation and documentation techniques used in the industry and identify challenges\\
	&\citeS{cerqueira2021exploring}	&Process&	This work is to offer a guide for eliciting ethical requirements in artificial intelligence projects (RE4AI Ethical Guide)\\
	&\citeS{ribeiro2022playful}&	Technique&	Developed and evaluated a Playful Probe protocol through a participatory design workshop, demonstrating how to elicit ideas for integrating maintenance planning practices with machine learning \\
	&\citeS{silva2019requirements}&	Model&	Present a new RE Model that allows software engineers and data scientists to discover these values hand in hand as part of the software requirement process\\
 
	&\citeS{nahar2022collaboration}&	Model&	Reveals key collaboration challenges and patterns in developing and deploying production ML systems, focusing on requirements, data, and integration\\
 
\midrule

Requirements Validation (8) 	&\citeS{challa2020faulty}&Technique&	Metamorphic testing for data quality requirements validation of DL systems \\
 
	&\citeS{singh2021}&	Technique&	Approaches to efficiently address the problem and derive component-level requirements and tests\\
	&\citeS{banks2019requirements}&	Model	&Develop nine specific areas where confidence is required in training data\\
	&\citeS{hu2020}&	Process&	Approach for specifying and testing requirements\\
	&\citeS{hu2022if}&	Model&	Comprehensive approach to enhancing MVC safety, including safety-related image transformations, reliability requirement classes, methods for creating machine-verifiable requirements\\
	&\citeS{barzamini2022cade}&	Technique&	A technique that evaluates datasets by bridging the gap between the specification of hard-to-specify domain concepts \\
	&\citeS{pradhan2023identifying}&	Process&	Presented includes a Data Quality Workflow, Lists of Data Quality Challenges and Attributes, and Solution Candidates, serving as tools for assessing and maintaining data quality, validated through a focus group\\
	&\citeS{riveiro2021s}&	Model&	Examining the system's output in scenarios that both align with and deviate from user expectations\\
\midrule

Practical Consideration (5) 	&\citeS{sheh2021explainable}&Model&	Categorisation of explanatory capabilities and requirements \\
	&\citeS{nakamichi2020requirements}&	Model&	Extending the quality characteristics of ISO 25010\\
	&\citeS{siebert2020}&	Model	&General methodological approach for quality Modelling of ML systems\\
	&\citeS{agbese2023ethical}&	Technique&	Method for applying ethical requirements in Agile portfolio management \\
	&\citeS{khan2022handling}&	Model&	Interaction between RE and Software Architecture in the context of ML\\
\midrule

Software Requirements Tool (4) 	&\citeS{henin2021}&	Tool&	Multi-layered approach allowing users to formulate their requests for explanations \\
	&\citeS{alagarswamy2023towards}&	Tool&	Implement Tool for exemplary audit requirements to demonstrate the applicability of a selected mobility application.\\
	&\citeS{agbese2023ethical}&	Tool&	Tutorial aims to teach SE stakeholders how to apply the Ethical Requirements Stack for implementing AI's ethical requirements across business levels, enhancing AI ethics research.\\
	&\citeS{shao2022data}&	Tool&	Tool support for Modelling Requirements\\

\bottomrule
\end{longtable}

\normalsize


\textbf{Requirements Analysis:}
This category deals with the comprehensive process of validating and managing stakeholders' needs and constraints to ensure a clear understanding and agreement on what the system or project must achieve. It includes activities such as conflict detection, prioritization, and scope definition.
The main themes found in requirements analysis were explainability and human-centric requirements analysis. $33$ papers proposed different practices in this area, with $5$ focused on requirement analysis for explainability needs. Sheh and Monteath \citeS{sheh2018} categorized explainability requirements while considering the source, depth, and scope of the explanation. Where \citeS{li2023dealing} introduces an explainability framework that automatically recommends methods to improve system design's explainability and efficiency for developers, thereby reducing the conflict between explainability and usability. K{\"o}hl et al. \citeS{kohl2019} provided a conceptual analysis that unifies the different concepts of explainability and the corresponding explainability demands. Suresh et al. \citeS{suresh2021} provided a framework for identifying stakeholders for interpretability and using the human cognitive process to derive requirements for explainability \citeS{wang2019}. Hall et al. \citeS{hall2019} outlined a systematic method to build an explainable artificial intelligence (XAI) system, which focuses on understanding specific explanation requirements and assessing existing explanation capabilities.
$4$ among $33$ papers were focused on human-centric requirements analysis, including human-centric design requirements \citeS{shergadwala2021}, or proposing RE frameworks that map the human mental model to ML models \citeS{camilli,habibullah2023exploring,heyn2023investigation} focused on ethical and legal requirement analysis. Other papers proposed requirement analysis for risk modeling
and frameworks for requirements analysis.

\textbf{Requirements Process:} This section of RE KA illustrates how the requirements process aligns with the overall SE process.
In the requirements process category, we find a paper proposing the process model of evidence-driven RE to capture the requirements specific to ML-based systems, i.e., uncertainty \citeS{ishikawa2020evidence}. 
Ries et at. \citeS{ries2021}, tailored traditional RE to improve dataset requirements engineering. At the same time, Vogelsang and Borg \citeS{vogelsang2019} highlighted; the need to integrate ML specifics in the RE process and new types of quality requirements such as explainability, freedom from discrimination, or specific legal requirements. Further, \citeS{habiba2022can} outlines challenges in Explainable AI (XAI) and proposes a framework for using RE practices to address these challenges.  
Similarly, \citeS{chuprina2021} proposed an artifact-based approach for the development of data-centric systems while \citeS{zhang2021dde} introduced a data-driven engineering process featuring hierarchical RE. \citeS{camilli2021} provided an overview of how research in the RE discipline can support building effective ML systems. Other authors proposed a modeling framework for analytics algorithms and data preparation activities \citeS{nalchigar2018,gauerhof2020assuring} or an agile data mining framework \citeS{qadadeh2020} in the context of business objectives.
However, another notable work \citeS{treacy2022legal} proposed a methodology for developing and assessing legal, privacy, social, and ethical requirements.

\textbf{Requirements Specification:} New specification practices have been proposed for AI-based systems to capture domain-specific and component-level requirements. Czarnecki and Salay \citeS{czarnecki2018} proposed an approach to specify safety requirements based on uncertainty, whereas Rahimi \citeS{rahimi2019} proposed an approach to specify requirements for ML components explicitly specifying domain-related concepts. Furthermore, ~\citeS{hu2020} and~\citeS{singh2021} focused on specifying requirements well suited to ML components and testing these requirements~\citeS{salay2019}.
Others focused on specifying requirements using user knowledge \citeS{gil2019towards} and providing the framework with a more granular and composable vocabulary to formulate user needs \citeS{suresh2021}.
Requirements Documentation and Evaluation has been highlighted by \citeS{vaida2021user} by providing a list of shared requirements. However, \citeS{eichelberger2022requirements} details usage view activities/scenarios, top-level requirements, and detailed sub-requirements. In this context, \citeS{bell2023think} proposes a transparency playbook for developing AI systems that meet legal, regulatory, and user requirements.
Furthermore, several studies have made contributions to defining specific requirements. For instance, \citeS{schuh2022case} extracts unique requirements from a case study, while \citeS{konigstorfer2021software} focuses on the necessities for AI documentation that bridge technical aspects with business processes. Berry and Daniel \citeS{berry2022requirements} highlight the importance of detailed evaluation measures and criteria in the specification process. Gr{\"u}ning \citeS{gruning2023towards} and Bartlett \citeS{bartlett2022characterizing} offer insights into the requirements for AI-enabled medical devices, with the former providing a comprehensive overview and the latter specifying sensor accuracy for stability scoring. Elshan et al.~\citeS{elshan2022requirements} delve into what is needed for an AI-based team member, and Noda \citeS{noda2021requirement} uses user personas from AI medical interviews to specify usability and reliability needs.

\textbf{Requirements Elicitation:} Elicitation considers the origin of requirements and how they can be gathered.
We identified among $16$, six papers proposing different models for requirements elicitation, of which four papers were focused on elicitation of explainability requirements \citeS{wolf2019,liao2020,jansen2021,chari2020,ahmad2023requirements}. The authors of \citeS{nakamichi2020requirements} proposed a method to identify requirements to ensure quality characteristics. Moreover, \citeS{siqueira2022guide} and \citeS{cerqueira2021exploring} provided a guide on how to elicit ethical requirements for AI-based systems. Further important requirement elicitation challenges related to data requirements are highlighted by \citeS{dey2023multi} and \citeS{nahar2022collaboration}.

\textbf{Requirements Validation:} 
While validating requirements is considered a crucial part of RE, we identified $8$ studies proposing new practices in this area. The Challa et al.~ \citeS{challa2020faulty} used a metamorphic testing approach to validate data quality requirements. Similarly, Banks and Ashmore \citeS{banks2019requirements} established that training data provides the functional requirements for AI-based systems. Using traditional assurance concepts, they developed nine areas where confidence is required in training data. Barzamini et al.~\citeS{barzamini2022cade} and Pradhan et al.~\citeS{pradhan2023identifying} presented a framework to evaluate data quality. However, \citeS{de2022requirements} suggested a model for examining the system's output in scenarios that both align with and deviate from user expectations.

\textbf{Practical Considerations:} The requirements process spans the whole software life cycle. This KA aimed to maintain stability in requirements to ensure they accurately reflect the software to be built or that has been built.
To support that, Sheh \citeS{sheh2021explainable} presents traceability between the explanations and the capabilities of underlying AI techniques to help users and developers. Authors in \citeS{nakamichi2020requirements} proposed a methodology to derive quality characteristics and measurement methods for MLS. Furthermore, a general methodological approach for quality modeling of ML has been proposed by \citeS{siebert2020}. Further, \citeS{agbese2023ethical} proposed a method to deal with ethical requirements, and \citeS{khan2022handling} addressed the interaction between RE and Software Architecture in the context of machine learning.

\textbf{Software Requirements Tool:} In total $4$ tools have been developed to address distinct needs. One such tool \citeS{henin2021} offers a multi-layered approach, enabling users to articulate their demands for explanations, facilitating a deeper understanding of AI systems. Another tool \citeS{alagarswamy2023towards} focuses on implementing tools for audit requirements, showcasing its utility with a mobility application to ensure compliance and functionality. Additionally, a tool~\cite{agbese2023ethical} has been created with the objective of educating SE stakeholders on utilizing the Ethical Requirements Stack, aiming to integrate AI's ethical requirements comprehensively and contribute to the advancement of AI ethics research. Lastly, there is tool support~\citeS{shao2022data} dedicated to modeling requirements, which assists in the precise definition and management of system requirements, underscoring the importance of clear and structured requirement specifications in successful system development.

\section{Open Challenges and Future Research Directions (RQ5)}
\label{sec:challenges}
This section highlights the prevailing challenges in the RE4AI literature and presents future research directions outlined among $126$ primary studies. We used the thematic synthesis approach recommended by Cruzes et al.~\cite{cruzes2011recommended} to answer challenges and future directions in RQ5.

\subsection{RE Challenges for AI-based Systems}
In this section, we underline the challenges in RE4AI. We identified $27$ challenges classified into $9$ categories as seen in Fig.~\ref{fig:challenges}. In the following subsections, we discuss them one by one. 

\subsubsection{Requirements Specification}

In the requirements specifications, we encountered the most challenges, categorized into five types:

\textbf{Hard to Specify Requirements Concretely}
The necessity for requirements engineers to adopt new methods to deal with data biases and the challenge of developing requirements when the data is not yet available highlight the difficulty in specifying requirements concretely for AI systems~\citeS{heyn2021}, \citeS{challa2020faulty}.
The challenge of ensuring that legal regulations and ethical considerations are adequately considered requires a shift in perspective towards a data and analytics viewpoint \citeS{vogelsang2019}.
The complexity of specifying non-functional requirements (NFR) on overall ML system performance and the difficulty of rigorously specifying requirements due to a lack of domain knowledge \citeS{vogelsang2019}, \citeS{hu2020}.
The challenge of specifying explainability requirements and functionality that depends on input data underscores the difficulty of concretely specifying requirements in AI-based systems~\citeS{kohl2019}, \citeS{salay2019}.
The difficulty of specifying unambiguous requirements, such as for a pedestrian detector component, further illustrates this challenge~\citeS{rahimi2019}.

\textbf{Incomplete and Incorrect Knowledge:}
Challenges around less tangible characteristics are hard to express meaningfully, leading to overlooked and misconstrued requirements~\citeS{sheh2021explainable}.
Incomplete, incorrect, and inconsistent knowledge encompassing missing or insufficient entities, mislabeled entities, and differing labels for the same entity or merged entities, highlighting issues of knowledge integrity~\citeS{maass2021}.

\textbf{Emergent Functionality Hard to Specify in Advance:}
The entanglement of requirements where even minor changes can dramatically alter other requirements illustrates the challenge of specifying emergent functionality in advance~\citeS{belani2019}.

\textbf{New Type of Quality Requirements:}
The explicit specification of explainability as a quality requirement presents a new challenge due to the lack of a systematic and overarching approach~\citeS{kohl2019}.
Standard requirements specification techniques become less applicable in AI-based systems where requirements are informed through training data, indicating a shift towards new types of quality requirements~\citeS{banks2019requirements}, \citeS{horkoff2019}.

\textbf{Lack of Suitable Guidelines for AI Documentation:} K{\"o}nigstorfer \citeS{konigstorfer2021software} and Treacy \citeS{treacy2022legal} underscore the issue of insufficient guidance on documenting AI, noting that many guidelines do not effectively connect principles with actionable requirements.

\subsubsection{Explainability Challenges}
Many studies have identified explainability as a noteworthy challenge. We arranged these challenges into three major categories:

\textbf{Explainability as a New Requirement:}
Ishikawa et al.~\citeS{ishikawa2020evidence} and Kuwajima et al.~\citeS{kuwajima2019} highlighted explainability as an emerging requirement, aligning with the European Commission's ethical guidelines for trustworthy AI, which advocate for fairness and explainability. This category underscores the recognition of explainability as a crucial aspect of ethical AI development.

  \begin{figure*}[htb]
  	\includegraphics[width=\textwidth,height=\textheight,keepaspectratio]{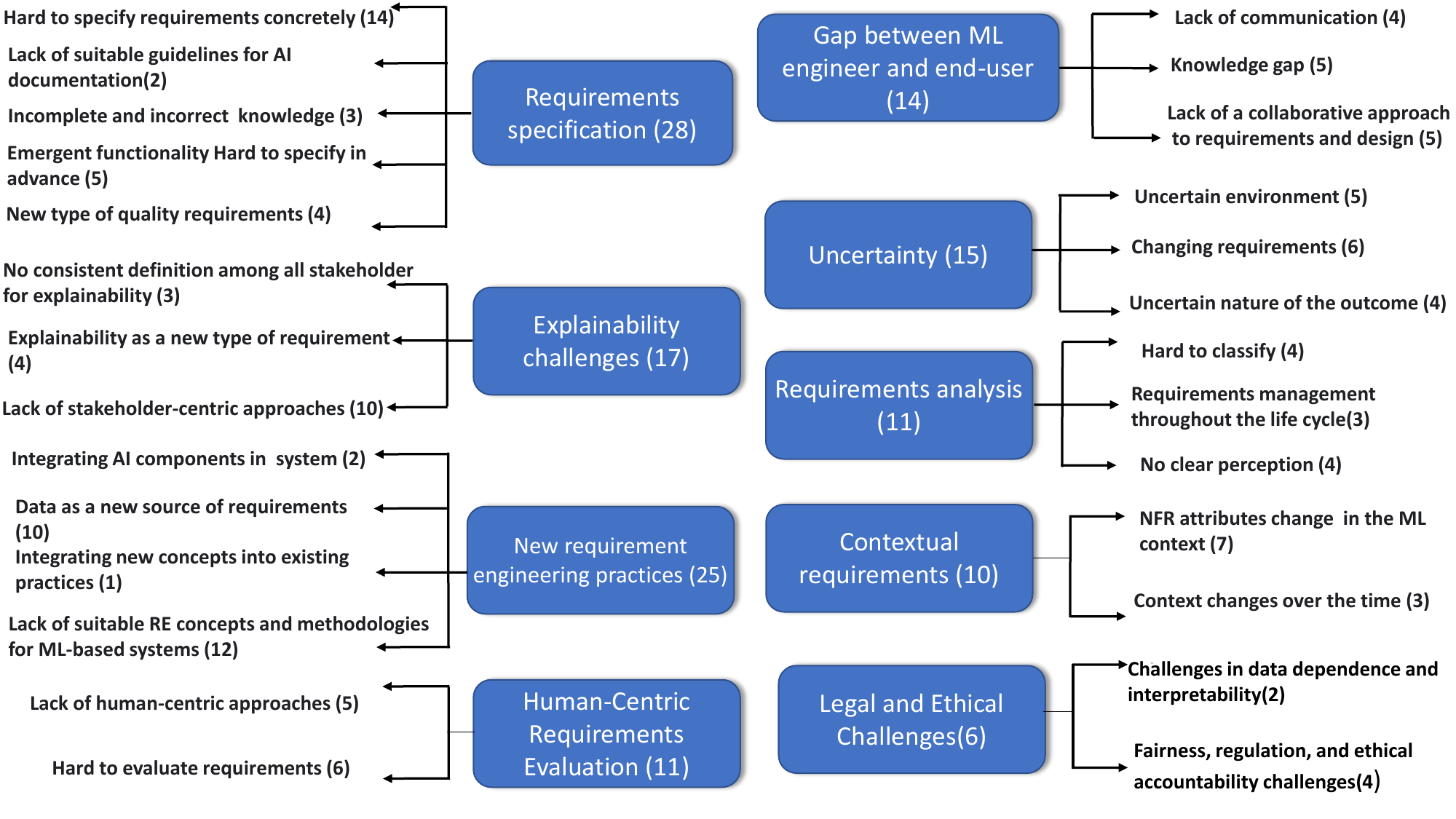}
  	\centering
  	\caption{Challenges Identified in Literature}
  	\label{fig:challenges}
  	
  \end{figure*}

\textbf{No Consistent Definition for Explainability:}
A significant challenge in the domain of explainability is the absence of a unified definition, making it difficult to pinpoint what 'explainability' precisely entails \citeS{jovanovic2022explainability}. This ambiguity is emphasized by studies like those of K{\"o}hl et al.~\citeS{kohl2019}, and Suresh et al.~\citeS{suresh2021}, who note that different stakeholders have varying interpretations of explainability. Furthermore, Jansen et al.~\citeS{jansen2021} and Kim et al.~\citeS{kim2023help} discuss the gap between stakeholders' expectations of AI explanations and their understanding of AI system actions, illustrating the complexity of achieving a common understanding of explainability across diverse groups.

\textbf{Lack of Stakeholder-Centric Approaches for Explainability:}
The necessity for stakeholder-centric approaches in explainability is underscored by the challenges in ensuring AI-based systems are transparent enough to foster trust and accountability \citeS{riveiro2021s}. Suresh et al.~\citeS{suresh2021} and Wang et al.~\citeS{wang2019} address the difficulties in creating AI-based systems that can effectively communicate their reasoning to users, particularly in critical situations. The literature suggests that existing model interpretability methods often fail to consider the end-user, typically being most comprehensible to those who develop them, such as ML researchers or developers \citeS{li2023dealing}. This point is further elaborated by Dhanorkar~\citeS{dhanorkar2021}, who argues for the need to extend beyond current explainability techniques to accommodate the diverse explanations required by different stakeholders in an AI system. Henin and Metayer~\citeS{henin2021} highlight the challenge of developing explanation methods that cater to various explainees with distinct interests, advocating for personalized approaches to explainability.

Collectively, these challenges indicate a growing awareness of the importance of explainability in AI, the need for a clearer definition and understanding of what explainability means to different stakeholders, and the importance of developing approaches that prioritize the perspectives and needs of those stakeholders

\subsubsection{New Requirements Engineering Practices}
The literature identifies critical areas where new Requirements Engineering (RE) practices are essential to address the unique challenges posed by AI-based systems. These areas are categorized into four key segments.

\textbf{Integrating AI Components in System:} The integration of AI components into systems presents novel challenges for RE, necessitating new validation techniques beyond traditional inspection and static reading, especially where data quality is paramount \citeS{challa2020faulty}. \citeS{kostova2020} highlights the need for a revised RE process pipeline to effectively address and evaluate the requirements for these AI components, underscoring the importance of safety, reliability, and effectiveness in AI systems \citeS{camilli2021}.

\textbf{Data as a New Source of Requirements:} Data quality and its role as a source of requirements for AI-based systems emerge as significant concerns. The traditional principles and techniques of RE are found inadequate in addressing the unique requirements of ML-based systems, prompting a reevaluation of existing RE practices \citeS{ishikawa2020evidence}.

\textbf{Integrating New Concepts into Existing Practices:} The challenge extends to integrating new concepts into established RE practices. Existing RE frameworks must evolve to accommodate the distinct needs of AI-based systems, requiring a comprehensive approach that includes strategic planning, technology selection, system validation, and maintenance processes \citeS{chuprina2021}.

\textbf{Lack of Suitable RE Concepts and Methodologies for ML-based Systems:} There is a conspicuous gap in RE concepts and methodologies tailored to ML-based systems. This deficiency points to a broader issue within the field, where RE practices fail to align with the legal and regulatory demands specific to ML systems. Ensuring compliance with relevant laws and regulations remains a primary concern for requirements engineers in this domain \citeS{vogelsang2019}.

\subsubsection{Human-Centric Requirements Evaluation}
\textbf{Lack of Human-Centric Approaches:}
The challenges across papers \citeS{habiba2022can,ahmad2023requirements,yu2022stakeholders,gruning2023towards,wang2022framework} collectively highlight a significant shortfall in human-centric approaches within AI system development. Habiba et al.~\citeS{habiba2022can} outline issues such as the lack of a mediator role for effective communication among stakeholders, the absence of a unified explainability definition, and the shortfall in stakeholder-focused development methodologies, alongside a missing common language for all involved in ML projects. These issues underscore a widespread neglect of human-centered perspectives in AI's technical evolution.

Ahmad et al.~\citeS{ahmad2023requirements} point out the increasing reliance on AI in software solutions that unfortunately often overlook essential human-centered considerations in favor of technical priorities, indicating a misalignment between technological progress and human values. Similarly, Yu and Yong~\citeS{yu2022stakeholders} expose a specific lack of engagement with the needs and perspectives of Korean stakeholders in AI for Health, revealing both a geographic and cultural oversight in stakeholder engagement. Gr{\"u}ning et al.~\citeS{gruning2023towards} discuss how companies frequently miss integrating user requirements in the innovation of business models and the creation of new AI products, especially in healthcare, leaving uncertain how AI might shape future business models in this vital sector. Lastly, Wang~\citeS{wang2022framework} criticizes the dominant focus on technical strategies like model extraction for interpretability, which neglects user expectations, highlighting a critical gap in aligning AI system development with actual user needs.

\textbf{Hard to Evaluate Requirements:}
Habibullah et al.~\citeS{habibullah2023non} underscore the importance of NFRs in maintaining ML system quality, noting differences in definitions and measurements of NFRs between traditional systems and ML systems, such as adaptability and maintainability. The difficulty in measuring NFRs like fairness and explainability due to their qualitative nature is compounded in safety-critical situations where both human and machine judgment are crucial. Additionally, challenges in NFR measurement are identified, including gaps in knowledge or practices, absence of measurement baselines, complex ecosystems, data quality issues, testing costs, bias in results, and domain dependencies. However, Bartlett~\citeS{bartlett2022characterizing} points out the complexity of defining sensor accuracy requirements to ensure reliable algorithm outputs, indicating a lack of straightforward or well-defined processes.
Similarly, Dey et al.~\citeS{dey2023multi} observe that while there is an emphasis on specifying ML-specific performance requirements, there is insufficient guidance on systematically engineering data requirements that involve diverse stakeholders.

\subsubsection{Gap between ML Engineers and End-Users}

This section focuses on the challenges arising from the gap between ML engineers and end-users. We categorized this gap into three distinct groups. 

\textbf{Lack of a Collaborative Approach to Requirements and Design:}
Initiating from a lack of collaboration,
Vogelsang and Borg \citeS{vogelsang2019} underlined that it is challenging for data scientists to explain performance measures and their relevance to the client in an effective and understandable way. Furthermore, to ensure that customers understand the performance measures, data scientists should also have skills in communication and customer education. Likewise, Shergadwala and El-Nasr \citeS{shergadwala2021} underscored the need to understand the shared mental model of design teams during human-AI collaboration. Liao et al. \citeS{liao2020} felt the need for explainability to make AI algorithms understandable to people. In contrast, Nalchigar and Yu \citeS{nalchigar2018} questioned the huge conceptual distance between business strategies, decision processes, and organizational performance. Lastly, Brennen \citeS{brennen2020} stressed it is essential to define a common terminology when discussing XAI to enable meaningful, productive conversations that can move the field forward. It could include establishing a shared vocabulary and clearly defined concepts and providing guidance on how to classify and rank models based on their explainability.

\textbf{Lack of Communication:}
Secondly, in lack of communication, a key challenge for software engineers developing ML systems is to determine how to capture customer requirements effectively and design user interfaces that effectively convey data to the user \citeS{rivero2021}. Similarly, another significant challenge is overblown expectations identified due to a lack of communication \citeS{wolf2019}.
To bridge user needs and technical capabilities to develop explainability systems that are flexible, responsive, and resilient to changing conditions is also a challenge \citeS{liao2020}. Qadadeh and Abdallah \citeS{qadadeh2020} stated that understanding the language and terminology used by data scientists and business users is a challenge in the context of data mining. They added that improving organizational communication between data miners and business analysts and finding a way to bridge the gap between theoretical research results in data mining and realistic project goals is also a challenge.

\textbf{Knowledge Gap:}
The third challenge is to build a shared understanding among stakeholders of the potential ML technology. Most importantly, it involves addressing the challenges in data collection and processing techniques, as well as implementing appropriate algorithms and models to ensure the overall effectiveness and reliability of the AI-based systems~\citeS{nalchigar2021,maass2021}.
Another challenge related to requirement elicitation for data analytics systems is to determine how to translate the business objectives into tangible and measurable analytics requirements. Additionally, there is a gap between non-technical stakeholders, who often have difficulty expressing their needs, and technical stakeholders, who need to understand and implement the requirements \citeS{nalchigar2018}.

\subsubsection{Uncertainty}
Uncertainty in AI-based systems presents significant challenges to the Requirements Engineering (RE) process, categorized into three distinct areas: uncertain environments, changing requirements, and the uncertain nature of outcomes.

\textbf{Uncertain Environment:} The uncertain environment encompasses challenges stemming from reliance on large volumes of data, where the accuracy and reliability of this data cannot always be assured \citeS{horkoff2019}. This situation is further complicated by unpredictable external conditions that might affect the system's performance and decision-making capabilities \citeS{camilli2021}, making it difficult to guarantee system behavior under varying conditions.

\textbf{Changing Requirements:} Changing requirements pose a persistent challenge, reflecting the dynamic nature of business and operational goals. As business requirements evolve, the technical side struggles to keep pace, especially in understanding and processing data effectively \citeS{nalchigar2021}. This fluidity can lead to discrepancies between expected and actual system capabilities, necessitating ongoing adjustments to the RE process.

\textbf{Uncertain Nature of the Outcome} The outcome's uncertain nature is particularly pronounced in AI-based systems, where the behavior on unseen data can significantly differ from expected results. This unpredictability complicates the RE process, as it undermines the ability to predict the system's performance accurately and, by extension, its development timeline, cost-effectiveness, and overall feasibility \citeS{ishikawa2020evidence,siebert2020,hu2020}. The inherent unpredictability of AI models demands a flexible and adaptive approach to requirements engineering, capable of accommodating unforeseen changes and outcomes.

\subsubsection{Requirements Analysis}

We identified three primary areas of concern in Requirements Analysis where each category reflects specific issues encountered in the development of AI-based systems, as delineated by primary studies.

\textbf{Hard to Classify:} The classification of requirements for AI-based systems into FR and NFR presents a significant challenge due to the inherent complexity of these systems. They leverage intricate algorithms and vast datasets, complicating predictions of system behavior in various scenarios or environments \citeS{belani2019}. The interaction of AI systems with external elements further amplifies this complexity, rendering traditional requirement analysis methods less effective. Moreover, the predictive nature of AI learning models complicates the advanced definition of system behavior, underscoring the classification challenge \citeS{nakamichi2020requirements}.

\textbf{Requirements Management Throughout the Life Cycle:} Effective requirements management across the lifecycle of an AI-based system is crucial yet challenging. \citeS{horkoff2019} emphasizes the importance of understanding the impacts of ML algorithms not only during the design phase but also post-deployment, advocating for a broader consideration of non-functional requirements (NFRs) beyond the integration of ML solutions. The non-deterministic behavior at runtime, influenced by the learning algorithms, complicates the classification and management of requirements for AI/ML-intensive systems \citeS{chuprina2021}. This dynamic behavior necessitates a flexible and adaptive approach to requirements management throughout the system's lifecycle.
 
\textbf{No Clear Perception:} 
"Clear perception" in the context of requirements analysis for ML-based systems refers to the precise and accurate understanding of how these systems perceive and interpret data from their environment. The lack of a clear perception during requirements analysis for ML-based systems poses a significant risk, potentially leading to the violation of other system requirements, such as data dependencies. This is particularly concerning for safety requirements in ML-intensive systems, where unclear or inaccurate perceptions can undermine the achievement of top-level safety goals \citeS{singh2021}. The challenge lies in adequately capturing and specifying these requirements in a manner that accounts for the nuanced and often unpredictable nature of ML-based perception.

\subsubsection{Contextual Requirements}

The concept of contextual requirements highlights the necessity of incorporating the specific environment or context in which an AI system operates into its design process. However, this presents two main challenges: accurately capturing and defining these contextual requirements, and integrating them into the design and development process. The variability and dynamic nature of real-world environments make it difficult to ensure that the AI system will perform optimally across different contexts. Traditional requirements engineering practices often fail to address these complexities, underscoring the need for new methods to effectively handle contextual requirements:

\textbf{NFR Attributes Change in the ML Context:} In the ML context, NFR attributes undergo significant transformations, necessitating a nuanced approach to their elicitation, specification, and validation. The complex nature of AI systems, coupled with the specific demands of their application domains, often results in a shift in the prioritization and characterization of NFRs. These shifts can be attributed to various factors, including emerging stakeholder expectations, evolving legal and ethical standards, and the technical requirements of integrating AI components. The challenge is compounded by a frequent lack of domain-specific knowledge, implicit stakeholder needs, and ill-defined problem scopes, making the accurate definition and management of these attributes particularly challenging \citeS{heyn2021,maass2021}.

\textbf{Context Changes Over Time:} Contexts within which AI systems operate are not static; they evolve over time, affecting the relevance and accuracy of the initially defined requirements. This dynamic nature of contexts can lead to significant alterations in requirement attributes, necessitating ongoing adjustments to both functional and non-functional requirements to maintain system efficacy and compliance. The ability to anticipate and adapt to these changes is crucial for the long-term success of AI systems, highlighting the need for flexible and responsive requirement engineering processes \citeS{horkoff2019}.

\subsubsection{Legal and Ethical Challenges}
\textbf{Challenges in Data Dependence and Interpretability:}
One significant challenge highlighted by Gabriel et al.~\citeS{gabriel2022requirements} is the reliance on extensive datasets and the domain expertise required to develop models while adhering to regulatory and ethical standards. There is a particular emphasis on the necessity to encapsulate implicit knowledge, especially from employees, and to ensure the AI system's operations are interpretable to them. The lack of practical experience with AI applications in many companies further complicates these challenges. Additionally, Silva et al.~\citeS{silva2022requirements} highlighted for ML systems, the inherent opacity poses a significant barrier to explainability, compounded by issues like ensuring non-discrimination, navigating legal restrictions on data usage, and the complex task of specifying data requirements.

\textbf{Fairness, Regulation, and Ethical Accountability Challenges:}
Treacy~\citeS{treacy2022legal} and Barclay~\citeS{barclay2021identifying} highlighted that current approaches lack mechanisms to extract protected attributes from legal requirements and to assist in the definition and interpretation of fairness in AI models, indicating a gap in developing fair AI systems. Further, Gr{\"u}ning~\citeS{gruning2023towards} stated companies aiming to offer AI solutions must navigate complex product requirements and complex regulatory landscapes, presenting significant operational challenges. Similarly, Cerqueira~\citeS{siqueira2022guide} emphasized that developers often lack adequate training in AI ethics, both in academic settings and within development projects. Furthermore, the absence of legal consequences for failing to implement ethical guidelines -- often because these guidelines are non-binding -- results in a lack of motivation or accountability among developers regarding AI ethics.

\subsection{Future Research Directions}
\label{futurework}
We propose future research in the directions of RE4AI, as outlined and summarized in Table~\ref{tab:future}. This proposal is founded upon a selective extraction of insights from primary studies.

\footnotesize 
 \begin{longtable}{>{\arraybackslash}p{1.6cm}>{\arraybackslash}p{9.5cm}}

 \caption{Future research directions}
\label{tab:future}\\
\toprule 

\textbf{\centering  Research Directions} &
 \textbf{ \centering  Descriptions {[}Paper ID{]}} \\ 

\midrule

\begin{nobullets} \item { Human in the loop}
\end{nobullets} &
\begin{itemize}  [topsep=0pt]
 \footnotesize \item A valuable research direction for software engineers is investigating how generative AI can enhance human understanding of domains or artifacts beyond just model outputs. \citeS{weisz2021}   

    \footnotesize \item Consider human factors to understand and ensure behavior and quality attributes of AL-intense systems. \citeS{heyn2021} 
 
    \footnotesize \item Creating design synthesis frameworks for computational tools that incorporate human-centric requirements \citeS{shergadwala2021}  
   \footnotesize \item Develop evaluation methods for AI systems focusing on usability from human and AI perspectives, through usability heuristics and AI scenario creation. \citeS{battistoni2023can} 
 \footnotesize \item Encourage empirical research on AI ethics within SE, focusing on practical solutions and integration into SE practices. \citeS{kemell2022utilizing} 
 
 \footnotesize \item Explore the importance of AI explanation in grasping user expectations, proposing the creation of user models for tailored explanations \citeS{riveiro2021s}
 
 \footnotesize \item Emphasize the value of considering stakeholders' sociotechnical backgrounds, historical information, and power structures in AI system development. \citeS{bell2023think} 
 \end{itemize} \\
 \hline

\begin{nobullets} \item {  Modeling AI requirements }
\end{nobullets}&

\begin{itemize}
\footnotesize \item Future research should address the challenges when extracting contextual definitions and requirements from use cases \citeS{heyn2021}

 \footnotesize \item New research directions for efficient top-down requirement derivation \citeS{singh2021} 
 
 \footnotesize \item Requirements modeling for ML-based systems. \citeS{yu2021} 

\footnotesize \item  Requirements modeling for Actor-with-Learning systems. \citeS{yu2021} 
 
 \footnotesize \item Examine the use of conceptual modeling to improve business understanding in the context of ML. \citeS{lukyanenko2019} 
\footnotesize \item Examine the use of conceptual modeling to improve the transparency, comprehensibility, and understandability of ML algorithm outputs \citeS{lukyanenko2019} 
 
\footnotesize \item Explore linking protected attributes via semantic analysis and user feedback, resolving conflicts among laws, and adapting to changing requirements. \citeS{lavalle2022law}
 
\footnotesize \item Enhance traceability with models detailing system, monitors, and their ties to high-level requirements like safety. \citeS{heyn2023investigation} 
 
\footnotesize \item Address the gap between mental/conceptual models and ML models by focusing on their interpretability and explainability for alignment. \citeS{maass2022conceptsuperimposition}

\end{itemize} \\

\hline

\begin{nobullets} \item {Applicability of  existing practices}
\end{nobullets} &

\begin{itemize}
 \footnotesize \item Investigate the suitability of GORE frameworks for integrating data-driven and model-based design (DMD) within the RE4AI taxonomy. \citeS{belani2019}

 \footnotesize \item  Using existing RE practices, e.g., GORE, for analysing the impact of ML on business goals. \citeS{nalchigar2021} 
 
 \footnotesize \item  How RE practices can support designing explainable system.~\citeS{kohl2019} 
 
\footnotesize \item   Assess the applicability of traditional trade-off analysis techniques in RE to AI contexts. \citeS{maalej2023tailoring} 

\footnotesize \item Develop a comprehensive quality engineering toolbox for AI, compiling techniques, tools, and guidelines for AI system development. \citeS{heck2023defining}

\footnotesize \item Utilize industry requirements as a benchmark for academic solutions, encouraging collaboration and real-world problem-solving in AI research. \citeS{bergelin2022industrial} 
\end{itemize} \\
\hline

\begin{nobullets} \item {RE process specific to AI}
\end{nobullets} &

\begin{itemize}
\footnotesize \item  Practices for specifying contextual requirements. \citeS{heyn2021} 

\footnotesize \item Developing suitable requirements information models, traceability strategies, and methodological support for ML-based systems. \citeS{heyn2021} 

\footnotesize \item   Exploring RE for ML systems. \citeS{vogelsang2019} 

\footnotesize \item  How can the RE for ML be incorporated with the RE for traditional software systems? \citeS{vogelsang2019} 

\footnotesize \item  Develop a method to deal with changing NFR. \citeS{horkoff2019}

\footnotesize \item  What could be a strategy to cope with uncertainty? \citeS{camilli2021} 

\footnotesize \item  RE would enable AI practitioners to systematically evaluate the actual use of AI in our system. \citeS{kostova2020} 

\footnotesize \item  RE practices for the systems that push their user to regress to the mean \citeS{kostova2020}

\footnotesize \item Adjust RE to emphasize data aspects, including collection, integration, preprocessing, labeling, and sharing in AI systems development. \citeS{maalej2023tailoring}

\footnotesize \item Develop a comprehensive catalog of roles and responsibilities in AI RE, incorporating agile interactions and emphasizing the need for REs to understand data characteristics. \citeS{franch2023requirements} 
 
\footnotesize \item Expand RE to cover additional scopes such as Data Engineering and Hardware requirements, ensuring a holistic approach to AI system development. \citeS{franch2023requirements} 

\footnotesize \item Advocate for a generalized approach to document and manage non-functional requirements (NFRs) in ML systems.\citeS{khan2022handling} 

\footnotesize \item Explore how traditional and new RE techniques can be applied to ML, utilizing perspective-based  task diagrams and specification templates.\citeS{villamizar2022towards}
 
\footnotesize \item  Develop RE methods to operationalize safety standards for training data selection in AI systems. \citeS{heyn2023investigation} 

\footnotesize \item  Address the balance between IP protection and safety argumentation information sharing through RE.\citeS{heyn2023investigation} 

\footnotesize \item  Create methods for identifying conditions suitable for runtime checks in AI systems.\citeS{heyn2023investigation} 

\footnotesize \item Formulate RE approaches to find runtime checks that maintain system performance without compromise.\citeS{heyn2023investigation} 
\end {itemize} \\
\hline

\begin{nobullets} \item {Non-Functional Requirements }
\end{nobullets} &
\begin {itemize} 
 \footnotesize \item  Managing and mitigating challenges related to data quality requirements. \citeS{heyn2021} 

 \footnotesize \item  Exploring synergies between data quality requirements and non-functional requirements. \citeS{heyn2021} 

\footnotesize \item   Exploring quality requirements related to MLS. \citeS{nakamichi2020requirements} 

\footnotesize \item   What are ML-specific requirements? \citeS{vogelsang2019} 

\footnotesize \item   Develop a comprehensive suite of NFRs to address general ML use cases. \citeS{horkoff2019} 
 
\footnotesize \item   Develop a language to specify NFR for ML-based systems. \citeS{horkoff2019} 

\footnotesize \item   Develop methods for quantifying and evaluating the trade-offs between competing NFRs for ML. \citeS{horkoff2019} 

\footnotesize \item   How to extract compliance requirements for ML-based systems. \citeS{camilli2021} 

\footnotesize \item Support the end-to-end provision of trustworthy information in ecosystems, requiring flexible, deployment-specific information exchange to meet diverse requirements.\citeS{barclay2021identifying} 

\footnotesize \item Emphasize research on NFRs of heightened importance in ML, such as explainability and bias, including new definitions and measurement methods.\citeS{habibullah2023non} 
 
\footnotesize \item Evaluate varying NFR importance across domains (medical, banking, automotive) and address misconceptions among practitioners and customers.\citeS{habibullah2023non}

\footnotesize \item Explore domain-specific NFR importance and conceptualize methods to categorize and define NFRs in ML systems. \citeS{habibullah2023non} 

\footnotesize \item Develop new NFR measurements tailored to ML, addressing the challenges of measurement in context-specific or domain-specific settings.\citeS{habibullah2023non} 
 
\footnotesize \item Advance the composition and relationship of quality models as a long-term goal, pushing for standardization in RE for AI.\citeS{franch2023requirements} 

 \footnotesize \item Transition from NFRs to quality requirements and constraints to align with current terminology and understand data limits. \citeS{franch2023requirements}

\footnotesize \item Investigate AI community concepts related to NFRs, such as data smells and model cards, for their integration into quality frameworks. \citeS{franch2023requirements}
\end {itemize} \\
\hline
\begin{nobullets} \item {Requiements Validation}
\end{nobullets} &

\begin {itemize} 
 \footnotesize \item Requirements validation for ML models. \citeS{heyn2021} 
 
 \footnotesize \item Validation of explainability requirements. \citeS{henin2021} 

\footnotesize \item Enhancing traceability of ethical requirements in code through evaluation methods and example mappings for AI systems \citeS{siqueira2022guide} 
 
\footnotesize \item Investigating transparency and explainability in AI through case studies, testing explainability requirements, and evaluating a model for defining such requirements.\citeS{balasubramaniam2023transparency}

 \footnotesize \item Create adaptive methods for evolving data and ML quality requirements, with triggers and regular reevaluations, informed by software evolution studies. \citeS{horkoff2019} 

\footnotesize \item Conducting empirical evaluations of models in big data projects to assess performance metrics and the business value of generated data.\citeS{altarturi2017requirement} 
\end {itemize} \\
\hline

\begin{nobullets} \item {New types of requirements}
\end{nobullets}  &

\begin {itemize} 
\footnotesize \item Considering new types of requirements, such as explainability, fairness, and freedom from discrimination. \citeS{dalpiaz2020} 
 
\footnotesize \item  DL-based software requirements also include dataset RE. \citeS{ries2021} 
 
\footnotesize \item How do we specify explainability requirements, what type of explainability is needed and who are the stakeholders for these explanations? \citeS{liao2020} 

\footnotesize \item How to specify requirements related to changes that cannot be expressed as transformations, e.g., changing clothes on a pedestrian? How to specify requirements related to changes that are not structure-preserving. \citeS{hu2020} 

\footnotesize \item Research to define a universal AI framework through examining AI measures, criteria, and contexts to ensure stakeholder needs are met. \citeS{berry2022requirements} 

\footnotesize \item Identify requirements for AI in healthcare focusing on laws, data use, reimbursement, and workforce development.\citeS{yu2022stakeholders} 

\footnotesize \item Investigate terminologies' effectiveness in AI projects for operationalizing responsible AI and enhancing requirements related to environmental impacts.\citeS{maalej2023tailoring}

\footnotesize \item Develop new methods for specifying data collection's variety and completeness, and improve ways to define context for data selection. \citeS{heyn2023investigation} 
\footnotesize \item Explore ethical implications of AI outcomes, suggesting progress in technology applications without clear ethical guidelines. \citeS{guizzardi2020ethical}

\footnotesize \item Propose a rigorous approach to understand and formalize requirements for superintelligence, including the ability to self-formulate goals. \citeS{kaindl2020towards}
\end {itemize} \\
\hline

\bottomrule
\end{longtable}
\normalsize


\textbf{RD1: How to incorporate human knowledge in building AI-system?}
New sophisticated and AI-enabled safety systems, such as automatic emergency braking (AEB), have dramatically transformed the relationship between human drivers and their respective cars. It frees up mental resources, enhances driving quality, and impacts other traffic participants and their conduct. While AI-powered driving assistance has evolved considerably recently, humans have remained the same over the previous millennia.
So, while building such features, we must consider several crucial factors (limitations and capabilities) from a human perspective. The fact that people may override or deactivate AEB capability, for example, has become a key constraint in its potential to make traffic safer. In this regard, considering to which extent human aspects must be included when examining the desired quality and needed functionality of the system and its components is a fruitful research opportunity \citeS{heyn2021}.
Also, how knowledge about human factors can be effectively incorporated into AI-intensive system development methodologies would be a promising research opportunity \citeS{shergadwala2021}. 

\textbf{RD2: How can requirements modeling be used for understanding AI-based systems?}
Requirements modeling enables the connection between domain problem understanding and technology solution, describing and justifying the step-by-step progression from problem to solution. Like conventional software, ML applications may benefit from well-known RE methodologies such as goal- and agent-oriented RE, ensuring that the final systems meet the goals and desires of end-users and other stakeholders \citeS{yu2021}. Furthermore, conceptual modeling can be seen as worthwhile to improve business understanding and enhance systems transparency \citeS{lukyanenko2019}. Where \citeS{singh2021} and \citeS{heyn2021} highlighted efficient top-down requirement formulation and deriving contextual requirements from use cases as a new research avenue.
 
\textbf{RD3: How can existing RE practices be adapted for AI-based systems?}
One of the major causes of poor ML system quality is the lack of requirements specification \cite{kuwajima2020engineering}. The main reason for this is the change in development paradigm and new types of requirements. \citeS{belani2019} and \citeS{nalchigar2021} outlined that research should be conducted to investigate how existing RE practices, e.g., GORE, data-driven and model-based design (MDM), can be adapted for AI-based systems.
The same applies to explainability: Kohl et al.~\citeS{kohl2019} emphasized that further research is needed to investigate how RE techniques can be applied to design explainable systems.

\textbf{RD4: How to identify the need for new RE practices specific to AI-based system?}
Several studies have shown that RE for ML systems is different because of the different ways these systems are developed; therefore, RE practices for these systems should also evolve. In this context \citeS{vogelsang2019} outlined the issues that future research should address: Is RE for ML distinct? If yes, what distinguishes it? If not, what are the reasons and consequences? Further research is needed to find how RE for ML can be integrated with the RE of a traditional software system \citeS{horkoff2019}.

\textbf{RD5: How to address non-functional requirements?}
A rigorous RE approach is required to assure quality. NFRs are requirements placed on system quality and are articulated over many quality characteristics \cite{habibullah2022non}. Further, the authors stated that our knowledge of NFR from the traditional system is no longer applicable to AI-based systems due to the non-deterministic behavior and additive performance requirements. One of the most critical aspects is data and its representation in ML systems since there needs to be an adequate mechanism to identify and manage the needed quality and amount of data \citeS{heyn2021}. Future research should focus on what are specific quality requirements related to ML systems, how these requirements can be specified \citeS{nakamichi2020requirements,horkoff2019}, particularly data quality requirements \citeS{heyn2021}, safety requirements \citeS{vogelsang2019}, and compliance requirements \citeS{camilli2021}.

\textbf{RD6: How to validate ML requirements?}
Some studies have dived into requirement validations for AI, but it is still in its infancy. Some possible future research directions are identifying appropriate performance metrics or key performance indicators (KPIs) for trained ML models in a particular context. Defining and monitoring the performance of ML systems ensures the system stays within its intended behavior \citeS{heyn2021}. In the context of requirements validation, \citeS{henin2021} pointed towards validating explanation as a potential future work. Some frameworks are provided to specify quality objectives as constraints or as criteria. Nonetheless, they need an evaluation of the relevance of such objectives, e.g., assessing the understanding of stakeholders, which can be a promising future research direction.

\textbf{RD7: How to address new types of requirements?}
This future research avenue should consider how to specify new types of requirements \citeS{dalpiaz2020,brunotte2022quo}? How to incorporate these requirements in the current development scenario \citeS{ries2021}? How do we formulate the specifications for alterations that defy expression as transformations, such as a person altering their attire, or when shadows obscure a portion of an object, which are not exemplified by the preservation of structure~\citeS{hu2020}?

\section{Discussions}
\label{sec:discussion}
This study systematically mapped the landscape of RE for AI-based systems, guided by five pivotal research questions. In this section, we will explore how both existing and new RE practices pose challenges for RE4AI and discuss future research directions to address these challenges. Our investigation reveals a multifaceted view of the current state, challenges, and future directions of RE practices in the domain of AI.

\textbf{RQ1 and RQ2: Landscape and Distribution of RE Practices}
According to our findings, academia has remained dominant throughout the years. However, the growing industry interest in RE4AI and the advent of industry contributions highlight the necessity for a specific RE process for AI-based systems development. Collaboration and industry contribution started appearing in 2019. Conferences and seminars that foster sharing of ideas and best practices might facilitate effective cooperation.

\textbf{RQ3: Maturity of Research}
Our analysis of RE publication types and evaluation methodologies found that more publications are proposing solutions, with the majority requiring additional industry validation.
This underscores the importance of additional research into evaluating AI-based system requirements as a significant opportunity for RE researchers. More research may be done to develop practices that help requirement engineers elicit the requirements of end users. Furthermore, new tool support for AI-based systems might be built to aid the RE process.
Research related to RE4AI is still relatively immature and dominated by proposals for solutions. Philosophical papers, opinion papers, and personal experience papers are fewer and have no evaluation. 
The research is primarily compelled by studies that validate the findings. Conducting experiments, case studies, and surveys can help researchers further develop the field.

\textbf{RQ4: Evolution of RE Practices}
According to our findings, there has been a strong emphasis on the requirements analysis and elicitation phases in RE4AI, with many existing and new approaches being used in these areas. However, the requirements process phase appears less mature than the other phases, particularly when adopting the requirements process for AI-based systems.
This is also notable no existing RE tool has been used to support RE4AI.


\begin{table*}[]
\footnotesize
\centering
\caption{Linking future directions, challenges and practices}
\label{tab:discussion}
\resizebox{\textwidth}{!}{%
\begin{tabular}{ll}
\hline
{\color[HTML]{000000} \textbf{Future directions}} & {\color[HTML]{000000} \textbf{Challenges {[}Paper ID{]}}} \\ \hline
 & \begin{tabular}[c]{@{}l@{}}Lack of human-centric \\ approaches \citeS{habiba2022can, ahmad2023requirements, yu2022stakeholders, gruning2023towards, wang2022framework}\end{tabular} \\ \cline{2-2} 
 & \begin{tabular}[c]{@{}l@{}}Lack of stakeholder-centric \\ approaches \citeS{riveiro2021s,suresh2021, wang2019, dhanorkar2021, li2023dealing, henin2021}\end{tabular} \\ \cline{2-2} 
 & \begin{tabular}[c]{@{}l@{}}No consistent definition among \\ all stakeholder for explainability \citeS{jovanovic2022explainability, kohl2019, suresh2021}\end{tabular} \\ \cline{2-2} 
 & Lack of communication \citeS{rivero2021, wolf2019, liao2020, qadadeh2020}\\ \cline{2-2} 
 & Knowledge gap \citeS{nalchigar2021, maass2021, nalchigar2018}\\ \cline{2-2} 
\multirow{-6}{*}{\begin{tabular}[c]{@{}l@{}}How to incorporate human \\ knowledge in building \\ AI-system?\end{tabular}} & \begin{tabular}[c]{@{}l@{}}Lack of a collaborative approach \\ to requirements and design \citeS{vogelsang2019, shergadwala2021, liao2020, nalchigar2018, brennen2020}\end{tabular} \\ \hline
 & \begin{tabular}[c]{@{}l@{}}Data as a new source of \\ requirements \citeS{ishikawa2020evidence}\end{tabular} \\ \cline{2-2} 
 & \begin{tabular}[c]{@{}l@{}}Lack of suitable guidelines for \\ AI documentation \citeS{vogelsang2019}\end{tabular} \\
\multirow{-3}{*}{\begin{tabular}[c]{@{}l@{}}How can requirements modeling \\ be used for understanding \\ AI-based systems?\end{tabular}} &  \\ \hline
 & \begin{tabular}[c]{@{}l@{}}Integrating new concepts into \\ existing practices \citeS{chuprina2021}\end{tabular} \\ \cline{2-2} 
\multirow{-2}{*}{\begin{tabular}[c]{@{}l@{}}How can existing RE practices \\ be adapted for AI-based systems\end{tabular}} & \begin{tabular}[c]{@{}l@{}}Integrating AI components \\ in system \citeS{challa2020faulty}\end{tabular} \\ \hline
 & \begin{tabular}[c]{@{}l@{}}Lack of suitable RE concepts and \\ methodologies for ML-based systems \citeS{vogelsang2019}\end{tabular} \\ \cline{2-2} 
 & \begin{tabular}[c]{@{}l@{}}Hard to specify requirements \\ concretely \citeS{heyn2021, challa2020faulty, vogelsang2019, hu2020, kohl2019, salay2019, rahimi2019}\end{tabular}  \\ \cline{2-2} 
 & Uncertain environment \citeS{horkoff2019, camilli2021}\\ \cline{2-2} 
 & Changing requirements \citeS{nalchigar2021} \\ \cline{2-2} 
\multirow{-5}{*}{\begin{tabular}[c]{@{}l@{}}How to identify the need for \\ new RE practices specific to \\ AI-based system?\end{tabular}} & Uncertain nature of the outcome \citeS{ishikawa2020evidence,siebert2020,hu2020}\\ \hline
 & \begin{tabular}[c]{@{}l@{}}NFR attributes change in\\  the ML context \citeS{heyn2021,maass2021}\end{tabular} \\ \cline{2-2} 
\multirow{-2}{*}{\begin{tabular}[c]{@{}l@{}}How to address non-functional \\ requirements?\end{tabular}} & New type of quality requirements \citeS{kohl2019, banks2019requirements, horkoff2019}\\ \hline
 & Hard to evaluate requirements \citeS{habibullah2023non, bartlett2022characterizing, dey2023multi}\\ \cline{2-2} 
 & \begin{tabular}[c]{@{}l@{}}Requirements management \\ throughout the life cycle \citeS{horkoff2019, chuprina2021}\end{tabular} \\ \cline{2-2} 
 & No clear perception \citeS{singh2021}\\ \cline{2-2} 
 & Hard to classify \citeS{belani2019, nakamichi2020requirements}\\ \cline{2-2} 
\multirow{-5}{*}{\begin{tabular}[c]{@{}l@{}}How to validate ML \\ requirements?\end{tabular}} & Context changes over the time \citeS{horkoff2019}\\ \hline
 & \begin{tabular}[c]{@{}l@{}}Explainability as a new type \\ of requirement \citeS{ishikawa2020evidence, kuwajima2019}\end{tabular} \\ \cline{2-2} 
 & \begin{tabular}[c]{@{}l@{}}Challenges in data dependence \\ and interpretability \citeS{gabriel2022requirements, silva2022requirements}\end{tabular} \\ \cline{2-2} 
 & \begin{tabular}[c]{@{}l@{}}Emergent functionality Hard \\ to specify in advance \citeS{belani2019}\end{tabular} \\ \cline{2-2} 
 & \begin{tabular}[c]{@{}l@{}}Fairness, regulation, and \\ ethical accountability challenges \citeS{treacy2022legal, barclay2021identifying, gruning2023towards, siqueira2022guide}\end{tabular} \\ \cline{2-2} 
\multirow{-5}{*}{\begin{tabular}[c]{@{}l@{}}How to address new types of \\ requirements?\end{tabular}} & \begin{tabular}[c]{@{}l@{}}Incomplete and incorrect \\ knowledge \citeS{sheh2021explainable, maass2021}\end{tabular} \\ \hline
\end{tabular}
}
\end{table*}
\normalsize
\textbf{RQ5: Challenges and Future Directions}
We identified nine main challenges and their subcategories that are frequently highlighted in the literature. We also extracted future research directions from selected primary studies during our analysis. The results indicate that there are still significant gaps, as illustrated in Fig.~\ref{fig:challenges} and opportunities (see Section~\ref{futurework}) to be addressed in RE practice for developing AI-based systems. 
Overall, RE4AI is still evolving and is a young field with the emergence of new types of AI-specific requirements. There is a strong need for specific RE processes for AI-based systems development, applying existing RE practices to AI-based systems, and addressing requirements specific to these systems.

Research, practitioners, and stakeholders in the field of RE for AI can benefit from our results as they allow them to understand the current state of research and identify areas where further research is needed. 
It provides a valuable resource for researchers planning to conduct future studies in this field by pointing out the challenges and research directions that still need to be explored. Furthermore, this study can be used as a reference for those who are planning to develop AI-based systems, as it provides an overview of the RE practices that are currently being used and those that are proposed for AI-based systems.

\textbf{Detailed Analysis of RQ4 and RQ5:}

In this section, we analyze RQ4 and RQ5, using Table \ref{tab:discussion} to link future research directions with identified challenges and relevant practices from Sections \ref{sec:existingpractices} (existing practices) and \ref{sec:newpractice} (new practices). This approach highlights how specific research directions address identified challenges and their correlations with old and new practices (see table \ref{Existing practices RQ2} \& \ref{tab:RQ2New}).

The studies emphasize the need for human-centric and stakeholder-centric approaches in RE4AI. For instance, Ahmad et al. \citeS{ahmad2023requirements} and Yu et al. \citeS{yu2022stakeholders} propose frameworks to collect human-centric requirements, addressing stakeholder engagement challenges. This aligns with new practices like those by Gruning \citeS{gruning2023towards} and Wang \citeS{wang2022framework} that enhance explainability. The future direction of incorporating human knowledge in building AI systems addresses the challenge of lacking human-centric approaches. Traditional structured methods for managing AI requirements can be adapted, and new practices enhancing explainability frameworks can be implemented.

Additionally, Ishikawa et al. \citeS{ishikawa2020evidence} introduce evidence-driven RE principles to manage the dynamic nature of data, linking goal-oriented RE for ML operations and hypothesis modeling. This adaptation of existing practices highlights the integration of new concepts into established frameworks, as suggested by Chuprina \citeS{chuprina2021}. The future direction of adapting existing RE practices for AI-based systems tackles the challenge of integrating new concepts into existing practices. Traditional goal-oriented RE methods can be combined with new practices like hypothesis modeling for continuous updates.

Furthermore, Heyn \citeS{heyn2021} and Maass \citeS{maass2021} discuss the evolving nature of NFRs in ML systems, linking these challenges to traditional quality requirements and new practices focusing on transparency and fairness. The future direction of addressing non-functional requirements in AI systems is crucial due to the changing NFR attributes in the ML context. Traditional quality requirements must evolve to include new quality requirements for transparency and fairness.

This analysis underscores the importance of adapting RE practices to meet AI-specific challenges, providing a comprehensive overview of how current research directions can address identified gaps and advance the field of RE4AI.

\section{Threats to Validity}
\label{sec: Threats to validity}
Like other secondary studies, our study is also prone to threats to validity~\cite{7890583}. We report them in the following sections along with the actions taken to mitigate them. We use guidelines provided by Petersen et al.~\cite{petersen2013worldviews} to assess the relevant threats to validity. As we adopt a pragmatist worldview, our study is prone to two main types of threats: (i) internal validity and (ii) external validity.

\subsection{Internal Validity}
Multiple factors can threaten internal validity.
For our study, the selected meta-search engines and digital libraries, the choice of articles, and the screening of articles are of particular interest. Internal validity threats can occur while developing search queries, choosing appropriate search engines, and applying inclusion and exclusion criteria. We piloted multiple search queries with different search engines. Then we selected the query with the most relevant results and the least noise. Two researchers independently ran the query on different search engines for this purpose. For the screening of articles, each article was reviewed by three researchers independently to reduce bias. Potential conflicts were resolved in a synchronization meeting. To avoid subjective bias during data extraction, we piloted the data extraction form and reviewed potential data extraction concerns.
To guarantee that all researchers followed a similar data extraction process, the first $15$ studies were independently extracted by the first three authors. We conducted weekly meetings with all researchers, during which, after each week, we discussed extractions for $5$ studies, after a week, the next $5$ studies, and so on. Further, the data presented can be reviewed by anyone who wishes to do so by accessing our replication package \footnote{See supplementary files for verification}, increasing the chance of identifying any errors in reporting (if any). Moreover, data analysis was conducted by a single researcher, the results were extensively discussed and refined by the research team.

\subsection{External Validity}
Threats to external validity are conditions that can affect the generalizability of our results. We ensured external validity by formulating an encompassing search query and applying rigorous inclusion and exclusion criteria. Although there is a chance of overlooking primary studies, we have tried to minimize this possibility. We identified the majority of relevant papers and applied snowballing until saturation, which resulted in a diverse set of articles that covered a significant and adequate part of the study topic. It helped us to increase the generalizability.

\section{Conclusion}
\label{sec:conclusion}
In this systematic mapping study, we presented a comprehensive overview of the research on RE in an AI-based system. 
We extracted 126 studies using a hybrid search strategy with iterative backward and forward snowballing and applying rigorous inclusion and exclusion criteria. We outlined the current research landscape based on our RQs among these studies, such as bibliometrics, including years, publishing venues, and author affiliations. We identified the contribution of the RE discipline towards AI and mapped RE topics addressed within RE4AI literature. 
Furthermore, the maturity of this research topic was evaluated using the RE publication classifications provided by Wieringa. We also highlighted how frequently these types occurred together and which evaluation methods were utilized. We also reported that existing RE practices have been used for AI-based systems and how many new practices have been proposed.
Moreover, we underscored seven significant challenges in the RE4AI and proposed seven potential future research directions based on our analysis.
Finally, the key contributions of this work are: (i) a mapping of the current state of research for RE4AI and (ii) the identification of challenges and prospective research avenues to further enhance RE4AI
Researchers and practitioners can use these results to familiarise themselves with the current state of this field.

\bmhead{Acknowledgments}
This work was partially supported by the German Federal Ministry of Education and Research in the projects KI B$^3$, grant number 21IV005E, and IKILeUS, grant number 16DHBKI041, and the Ministry of Science, Research and Arts Baden-W{\"u}rttemberg in the Artificial Intelligence Software Academy (AISA).

\bibliography{sn-bibliography}
\bibliographystyleS{sn-mathphys}
\bibliographyS{referenceA}


\end{document}